\documentclass[fleqn,usenatbib]{mnras}

\usepackage{newtxtext,newtxmath}

\usepackage[T1]{fontenc}

\DeclareRobustCommand{\VAN}[3]{#2}
\let\VANthebibliography\thebibliography
\def\thebibliography{\DeclareRobustCommand{\VAN}[3]{##3}\VANthebibliography}

\usepackage{graphicx}

\usepackage{babel}

\usepackage{amsfonts}
\usepackage{ulem}

\newcounter{tr}
\usepackage[dvipsnames]{xcolor}

\usepackage{amsmath}
\ifnum \value{tr}>5

\newcommand{\deleted}[1]{{\color{red} Damien - Deleted: } \sout{#1}}

\newcommand{\authorcomment}[1]{{\color{red} Damien - Comment :} {\color{Cyan} #1}}

\else

\newcommand{\deleted}[1]{}

\newcommand{\authorcomment}[1]{}

\fi
\newcommand{\txs}{TXS 0506+056}
\newcommand{\source}{3C 279}
\newcommand{\hsp}{3HSP J095507.9+355101}
\newcommand{\gray}{$\gamma$-ray}
\newcommand{\grays}{$\gamma$-rays}

\title[Simulator  of  Processes  in Relativistic AstroNomical Objects]{Time-dependent lepto-hadronic modeling of the emission from blazar jets with \textit{SOPRANO}: the case of \txs{}, \hsp{} and \source{}}

\author[Gasparyan, B\'egu\'e, Sahakyan]{S. Gasparyan$^{1}$\thanks{Corresponding author : sargisgyan@gmail.com }, D. B\'egu\'e$^{2,3}$ \thanks{E-mail: cayley38@gmail.com} \& N. Sahakyan$^{1,4,5}$ \thanks{E-mail: narek@icra.it} \\
$^1$ ICRANet-Armenia, Marshall Baghramian Avenue 24a, Yerevan 0019, Armenia\\
$^2$ Department of Physics, Bar Ilan University, Ramat-Gan 52900, Israel \thanks{Current address.}\\
$^3$ Max-Planck-Institut f\"ur extraterrestrische Physik, Giessenbachstrasse, D-85748 Garching, Germany\\
$^4$ ICRANet, P.zza della Repubblica 10, 65122 Pescara, Italy\\
$^5$ ICRA, Dipartimento di Fisica, Sapienza Universita` di Roma, P.le Aldo Moro 5, 00185 Rome, Italy
}

\date{Accepted XXX. Received YYY; in original form ZZZ}

\begin{document}
\label{firstpage}
\pagerange{\pageref{firstpage}--\pageref{lastpage}}
\maketitle


\maketitle

\begin{abstract}
    The observation of a very-high-energy neutrino by IceCube (IceCube-170922A) and its
    association with the flaring blazar TXS 0506+056 provided the first multimessenger
    observations of blazar jets, demonstrating the important role of protons in their dynamics and emission. In this paper, we present \textit{SOPRANO} (\url{https://www.amsdc.am/soprano}), a new conservative implicit kinetic code which follows the time evolution of the isotropic distribution functions of protons, neutrons and the secondaries produced in photo-pion and photo-pair interactions, alongside with the
    evolution of photon and electron/positron distribution functions. \textit{SOPRANO} is designed
    to study leptonic and hadronic processes in relativistic sources such as blazars and gamma-ray
    bursts. Here, we use \textit{SOPRANO} to model the broadband spectrum of \txs{} and \hsp{},
    which are associated with neutrino events, and of the extreme flaring blazar \source. The
    SEDs are interpreted within the guise of both a hadronic and a hybrid model. We discuss the
    implications of our assumptions in terms of jet power and neutrino flux.
\end{abstract}

\section{Introduction}

The discovery of the first cosmic very high energy (VHE; $>100$ GeV) neutrinos in 2013 by the IceCube experiment
\citep{Ice13,AAA13,AAA20} has opened a new window on VHE sources such as gamma-ray bursts (hereafter GRBs), active
galactic nuclei (AGNs) and tidal disruption events (TDEs). The lack of high confidence association between these
neutrino events and a particular type of sources significantly complicated the interpretation of their origin.
Potentially, they are produced in the sources where ultra-high energy cosmic rays (protons or nucleons with
energy exceeding $10^{19}\:{\rm eV}$) are accelerated. If the origin of these neutrinos remains an open question,
the VHE neutrino event IceCube 170922A \citep{IceCube18} and its $3.5\sigma$ association with the (simultaneously)
flaring blazar TXS 0506+056 \citep{IceCube18b, 2018MNRAS.480..192P} made clear that high energy (HE; $>100$ MeV) protons, neutrons and even possibly
nucleons have an important role to play in the dynamics and the radiation of relativistic jets \citep{BRS90,SSB96,AD03}.

Blazars are a subclass of AGNs which have their jet  aligned with or making a small angle to
the observer \citep{1995PASP..107..803U}. Blazars are among the most luminous and energetic
sources in the Universe. Based on optical emission lines blazars are sub-grouped as flat spectrum radio
quasars (FSRQs) and BL Lacs: the emission lines are strong and quasar-like in FSRQs and weak or absent in BL
lacs \citep{1995PASP..107..803U}. The emission from blazar jets, extending from radio to HE and VHE
\gray{} bands \citep{2017A&ARv..25....2P}, is characterized by rapid and high amplitude
variability, especially in the HE and VHE \gray{} bands \citep[e.g., ][]{2014Sci...346.1080A,2016ApJ...824L..20A}.
This variability suggests that the emission originates from a compact relativistically moving region. Since the
\gray\ emission has been detected even from blazars at very high redshift, $z>3.1$, \citep[e.g., ][]{ 2016ApJ...825...74P, 2017ApJ...837L...5A, 2020MNRAS.498.2594S}, they are unique objects to study the
evolution of jet power, morphology and emission processes in different cosmic epochs.

The broadband spectral energy distribution (SED) of blazars typically exhibit a double hump
distribution, the first one peaking at optical/UV or X-ray bands (low energy component) and the other one
in the HE or VHE \gray\ bands (HE component). The low energy component is usually explained by
synchrotron radiation of relativistic electrons in the jet magnetic
field. The origin of the HE component is still under
debate, mostly between two main scenarios. In leptonic scenarios, the HE component is due
to inverse Compton scattering of low energy seed photons by the relativistic electrons in the blazar jet
\citep{ghisellini, 1992ApJ...397L...5M, bloom}. The nature of the seed photons depends on the
location of the emission region and can be produced either inside or outside
the jet \citep[e.g., ][]{2009ApJ...704...38}. In the alternative hadronic scenarios,
synchrotron radiation from protons, see \textit{e.g.} \citet{2001APh....15..121M}, and
secondaries generated in photo-pion and photo-pair interactions produce the
emission from the X-ray to the HE \grays{} bands \citep{1993A&A...269...67M, 1989A&A...221..211M, mucke2}. 
In addition, inelastic $pp$ scattering could be involved when the highly energetic protons of the jet
interact with a dense proton target, such as clouds in the broad line region or surrounding stars (\textit{e.g.}
\citet{1997ApJ...478L...5D,1999ApJ...510..188B,2013MNRAS.436.3626A}).

Protons are unavoidably accelerated with the electrons in the jet,
but a direct test of their presence and energy cannot be done when only considering
electromagnetic data. Except for the cases when the
leptonic models face severe problems to account for the observed data,
usually both leptonic and hadronic models give equally good representation of the data \citep[e.g., ][]{2013ApJ...768...54B}. Indirect test of proton content and a
proof of the hadronic origin of the HE and VHE emission can
only be given by the observation of VHE neutrinos. Indeed, when protons
interact within the jet, the energy they lose is nearly equally divided
into electromagnetic and neutrinos components.
The produced neutrinos escape the emitting region, carrying information
about the protons in the jet and their distribution function.

Multimessenger observations have long been considered the next major breakthrough required for the study of extra-galactic objects. The recent  association of IceCube 170922A \citep{IceCube18} with \txs\ provided the
first ever possibility to perform a direct multimessenger study of a blazar jet. 
In addition, an analysis of the IceCube archival data revealed a $\sim13$ neutrinos excess within a 110 day period,
between September 2014 and March 2015, in the direction of \txs{}. Those two pieces of information together suggests that \txs\ is indeed the source of those HE neutrinos.  Moreover, a second possible
association between the muon track event IceCube 200107A \citep{2020GCN.26655....1I} and the blazar \hsp{} in
a flaring state was reported based on the small angular distance ($0.62^\circ$) between \hsp{} and the best-fit
position of IceCube 200107A \citep{2020A&A...640L...4G,2020ApJ...902...29P}. These two associations provide
unprecedented data allowing to constrain the hadronic processes in relativistic jets. 

In order to exploit multiwavelength and multimessenger data-sets, several groups
have developed numerical models to estimate leptons, hadron and photon distribution functions, either
under the steady state approximation \citep{AI07,BRS13,CZB15,ZCM17}, or in a fully time dependent approach
\citep{MC95,PW05,BMM08,VP09,DMP12,DBF15, GPW17, KMB20, Jv21} and use them to model the broadband SED of blazars
and other relativistic transients. Time-dependent modeling of blazars, both leptonic and hadronic is required to
understand the time evolution of particle distribution functions, for instance during a flare, see
\textit{e.g.} \cite{BB19}.

Time-dependent hadronic modeling is challenging as many different particles are involved. The time evolution of the initial particle populations, as well as that of the secondaries, should be treated with a set of kinetic equations, where
the energy is conserved in a self-consistent manner, i.e., the energy lost by a particle is exactly transferred to the energy of other particles. In this paper, we present and use a new fully time-dependent hadronic code,
\textit{SOPRANO}\footnote{\url{https://www.amsdc.am/soprano}}, standing for Simulator of Processes in Relativistic AstroNomical Objects, which takes into account all
relevant processes (leptonic and hadronic) and allows to compute the SED in any given period. The code solves the
time dependent isotropic kinetic equations and preserves the total energy of the system as well as the number of
particles where needed. The code structure is modular such that processes can be easily added (or removed). \textit{SOPRANO} is
implicit so numerical stability is achieved at all time. The code is designed in a such manner that by changing the initial
conditions, the lepto-hadronic processes can be investigated in blazar jets, GRBs and other
relativistic astrophysical sources where protons are hypothesized to be efficiently accelerated.

The paper is organised as follow. Section \ref{sec:intro_soprano} gives a short description of
our kinetic code \textit{SOPRANO}. The kinetic processes included in our numerical code and their
cross-sections are detailed in Appendix \ref{sec:kinetic_equation}. The numerical discretization in
energy and in time is provided in Appendix \ref{sec:numerical_discretization}. The analytical estimates
of several key model parameters are provided in Section \ref{analit_est} whereas the code
is applied to model the broadband SEDs of \txs{}, \hsp{} and \source{} in Section \ref{sec:applications_TSX}.
The discussion is in Section \ref{disc}, whereas the conclusion is summarized in Section \ref{conc}. Throughout the paper, we use the definition
$X = X_x \times 10^x$ where a quantity $X$ is given in cgs units. Moreover, the following cosmological
constants are adopted: $\Omega_M$ = 0.3, $\Omega_\Lambda$ = 0.7, and $H_0$ = 70 km $s^{-1}$ ${\rm Mpc}^{-1}$ \citep{2001ApJ...553...47F}.

\section{\textit{SOPRANO}: Simulator Of Processes in Relativistic AstroNomical Objects}
\label{sec:intro_soprano}

Investigation of hadronic processes in galactic sources, such as supernovae remnants and pulsar wind nebulae,
as well as extra-galactic objects, such as AGNs and GRBs, has always been an interesting but challenging task. Primarily, it is related with the desire to identify the sources in which cosmic rays and ultra-high energy cosmic rays are accelerated, and to understand the processes responsible for the broadband emission. Such studies are especially timely after the recent IceCube observations of cosmic neutrinos and their association with blazars. Indeed, for the first time, it is possible to constrain the emission process using a different window than that of electromagnetic observations.

In order to interpret the observed data and constrain the models that can explain the observed VHE neutrinos, it is necessary to perform self-consistent simulations of the time evolution of the distribution functions of all interacting particles: protons, neutrons, photons, electrons and positrons, as well as of the secondaries produced in photohadronic
interactions, such as pions, muons and neutrinos. This is a challenging task since {\it i)} there is a large number of distribution functions (fourteen even though
some are trivial), {\it ii)} all equations describing the time evolution of particle distribution functions are
coupled in a non-trivial and non-linear way by many complex processes that {\it iii)} have very different time scales,
requiring an implicit time discretization. 
The high number of distribution functions is necessary to compute the cooling and emission of charged secondaries, pions and muons. This requirement also prevents the use of semi-analytical expressions for the production rate of neutrinos, as given in \textit{e.g.} \cite{KA09}
For blazars, the magnetic field is expected
to be around or smaller than $1$G for leptonic models \citep[see \textit{e.g.}][]{FDB08,GT09,TG16,GSB18}, while
hadronic models usually require the magnetic field to be larger, in the range of few tens to few hundreds
Gauss \citep{RMR11,ZCM17}, see however \citet{KT17}. This magnetic field is too low to observe a substantial modification of the neutrino spectrum \citep{BT20}. However, synchrotron cooling of secondaries
produces photons, which form pairs, which in turn will radiate, effectively shifting the spectrum to lower-energies
for which strong constraints are given by X-ray observatories. In fact, X-ray observations are believed to be the most
constraining ones for hadronic models of blazars. In particular, they strongly challenge any models attempting
to explain the neutrino emission of TSX 0506+056 \citep{KMP18,CZB18,GFW19, XLP19}.

With the goal to model the multiwavelength and multimessenger SED of relativistic sources (e.g., AGNs and GRBs),
we have developed a numerical code which computes the temporal evolution of particle distribution functions by
solving the relevant kinetic equations. This code, \textit{SOPRANO}, relies on two underlying assumptions : \textit{i)} the space is homogeneous and \textit{ii)} particle distribution functions are isotropic. In its current version, \textit{SOPRANO} uses implicit time discretization to evolve the distribution functions of the following particles: 
\begin{enumerate}
\item photons,
\item electrons and positrons, considered as a single particle type,
\item protons,
\item neutrons,
\item charged ($\pi^+$, $\pi^-$) and neutral pions ($\pi^0$) separately
\item muons
\item neutrinos and anti-neutrinos, all species separately. 
\end{enumerate}
The processes considered for the above listed particles are: 
\begin{enumerate}
\item synchrotron emission and cooling of all charged particles (protons, electrons and positrons, charged pions and muons),
\item inverse Compton scattering of photons by electrons and positrons,
\item Bethe-Heitler photo-pair production and corresponding proton cooling,
\item photo-pion production and corresponding cooling of protons and neutrons,
\item pion and muon decay,
\item neutrino production.
\end{enumerate}
Detailed expression for the interactions kernel and all terms appearing in the kinetic equations for all particle species are given in Appendix \ref{sec:kinetic_equation}.

The energy discretization of the fourteen coupled partial differential equations is presented in Appendix
\ref{sec:numerical_discretization}. It follows from the prescription of finite volume allowing us to conserve particle number to machine accuracy for all processes which conserve particle number. For instance, for pion decay, there are as many muons and neutrinos created as pions that decay. Our numerical implementation ensures
that $\partial n_\pi/\partial t = -\partial n_\mu/\partial t = -\partial n_\nu/\partial t$. Energy conservation is also enforced by specific choices for the fluxes for diffusion-like terms or redistribution of particles between adjacent energy cells. The difficulty in our numerical implementation is in the computation of the 3- to 5-dimensional integrals which approximate the rates on each energy bin. Each of those integrals are computed to a relative accuracy of $10^{-4}$ with locally adaptive Gauss-Kronrod method. They only need to be computed one time for a given grid and since we do not change the energy grid, they remain the same for all the results presented here.

The largely varying time-scale of the processes and the large energy span of particle and photon grids require using an implicit scheme for the time integration. The code uses a semi-implicit version of the backward Euler method, that is to say that for the evaluation of photo-pion and photo-pair collisional terms, the photon spectrum is assumed to be explicit, while the proton and neutron distribution functions are solved for implicitly. This assumption makes the kinetic equation for all hadrons linear by decoupling their evolution from that of the photons and pairs. In practice, it means that the rate of photo-pair and photo-pion interactions might be underestimated, unless the time step is carefully chosen. We have studied how the time step of the integration method should be chosen to minimise the impact on the solution. Then, the kinetic equations describing the evolution of leptons are solved fully implicitly. The product terms $n_{\rm ph} n_{\rm e}$ and $n_{\rm ph} n_{\rm ph} $ appearing in Compton scattering and pair production make the problem non-linear and the coupled kinetic equations are solved with the Newton-Raphson method. We have checked that our code is able to properly account for particle cooling as well as to reproduce semi-analytical examples. Those tests are presented in Appendix \ref{sec:code_tests}. 

\section{Model Setup: Analytical estimation of model parameters}

\label{analit_est}
The broadband spectrum of blazars extends from radio to the HE or VHE \gray\ bands, covering a large
$10^{20}$ Hz frequency range \citep[e.g., ][]{2017A&ARv..25....2P}. The observed nonthermal emission
is produced in the jet and can be explained by different models. The primary dichotomy is the split between
leptonic and hadronic models, depending on the type of particles (electron-positron pairs or hadrons) initiating
the emission. On the one hand, leptonic models are solely based on the synchrotron emission of relativistic electrons
at low energy, while the HE peak is explained either by synchrotron self-Compton, hereinafter SSC, or by external
Compton process. These models assume that proton emission has a negligible
contribution to the overall SED, and therefore lack the ability to produce a significant amount of
VHE neutrinos ($\sim 10^{15}$eV) as detected by the IceCube observatory \citep{Ice13,AAA13}. In contrast,
the so-called hadronic models assume that protons are also efficiently accelerated in the jet and contribute
to the multiwavelength spectrum either by the synchrotron process, or by the radiation from the secondaries
produced in photo-pair and photo-pion interactions.

The modeling of the observed SEDs, be it leptonic, hadronic or lepto-hadronic, is a regular approach and
is a unique way to investigate the physical processes taking place in jets. 
The particle spectra are defined by the acceleration and cooling processes within the jet, which may vary
from source to source. In this work, we assume that particles are instantaneously accelerated and injected
in the emission zone where they radiate their energy. The particle injection spectrum is usually assumed to be a simple power-law, a power-law with an exponential cutoff or a broken power-law. Additionally, the emitting region can contain broad
external photon fields which interact with the relativistic particles in the jet. For instance,
photons emitted by the dusty torus or reflected by the broad line region play a crucial role in shaping
the multiwavelength emission of FSRQs \citep[e.g., order of minutes, ][]{ 2009ApJ...704...38,SBR94,GT09}. Moreover an arbitrary distributed photon field can be
considered as well, which is necessary for complex scenarios such as the multi-zones or the spine-sheath
layer models \citep{2008MNRAS.385L..98T}. \textit{SOPRANO} is designed to work with arbitrary injection
particle spectrum as well as arbitrary external photon field, and proceed to compute the evolution of
particle spectrum. This makes \textit{SOPRANO} an ideal code to investigate the emission processes in
different astrophysical environment.

Within the leptonic and hadronic interpretation of the blazar SEDs, it is assumed that the
emission is produced in a spherical blob of comoving size $R^{\prime}$ that moves towards the observer with a bulk
Lorentz factor $\Gamma \sim \delta$, where $\delta$ is the Doppler factor. Accelerated leptons
and hadrons are injected in the emitting region, which is uniformly filled with a magnetic field of
strength $B$. The magnetic jet luminosity is 
\begin{align}
  L_{\rm B} =  \pi c R^{\prime 2} \delta^2 \frac{B^2}{8 \pi} \label{eq:Lb_jet},
\end{align}
where $c$ is the speed of light. We assume that protons are injected in
the comoving frame with a power-law spectrum:
\begin{align}
Q^{\prime}_{\rm p}(\gamma_{\rm p})=Q^{\prime}_{0,p} \gamma_{\rm p}^{- \alpha_{\rm p}} & ~~~~~~~ & \gamma_{\rm p} < \gamma_{\rm p, max}.\
\label{pdist}
\end{align}
The normalization factor $Q^{\prime}_{0,p}$ is linked to the proton luminosity as
\begin{align}
L_{\rm p}= \pi R^{\prime 2} \delta^2  m_{\rm p} c^3 \int \gamma_{\rm p} Q^{\prime}_{\rm p}(\gamma_{\rm p}) \gamma_{\rm p}, \label{eq:Lp_jet}
\end{align}
where $m_{\rm p}$ is the proton mass. We assume that the injection electron spectrum is given by a power-law
with an exponential cut-off :
\begin{equation}
Q^{\prime}_{\rm e}(\gamma_{\rm e})= \left \{ \begin{aligned}
& Q^{\prime}_{0} \gamma^{-\alpha_{\rm e}}_{\rm e} \exp \left ( - \frac{\gamma_{\rm e}}{\gamma_{\rm e, cut}} \right ) & ~~~~ & \gamma_{\rm e, min} \leq \gamma_{\rm e}  \leq \gamma_{\rm e, max},  \label{edist} \\
& 0 & & {\rm otherwise,}
\end{aligned} \right.
\end{equation}
where $\gamma_{\rm e, min}$ is the minimum injection Lorentz factor.
The electron luminosity is then given by 
\begin{align}
L_{\rm e}=\pi R^{\prime 2} \delta^2  m_{\rm e} c^3 \int \gamma_{\rm e}\:Q^{\prime}_{\rm e}(\gamma_{\rm e})\:d \gamma_{\rm e} , \label{eq:Le_jet}
\end{align}
where $m_{\rm e}$ is the electron mass. In general $\gamma_{\rm e, cut}$ should be defined by the equality of the acceleration and
cooling time-scales. However, in order to have a broad inference of the physical processes in the
jet, $\gamma_{\rm e, cut}$ is considered a free parameter which will be constrained by the data.
The distribution functions of protons and electrons evolve via cooling and via interaction with photons, producing
different signatures in the broadband spectrum. Our aim is to identify those signatures and use them to constrain
the emission mechanism within the framework of different scenarios.

In one dynamical time-scale, $t_{\rm d}^\prime \sim R^{\prime}/c$,
electrons and positrons cool to Lorentz factor 
\begin{align}
    \gamma_{\rm e, c} = \frac{6 \pi m_e c^2}{B^2 R^{\prime} \sigma_{\rm T}} \sim 2.3 \times10^{3}  B^{-2} R_{16}^{\prime -1},
    \label{eq:gamma_c_pm}
\end{align}
where $\sigma_{\rm T}$ is the Thompson cross-section. The associated observed synchrotron characteristic frequency is 
\begin{align}
    \nu_{\rm e, c} = \frac{48 \pi \delta c^3 m_{\rm e} q}{B^3 R^{\prime 2} \sigma_{\rm T}^2} \sim 4.0 \times 10^{14}  \delta_1 B^{-3} R_{16}^{\prime -2} ~~~ {\rm Hz,}
    \label{sync:peak}
\end{align}
where $q$ is the elementary charge.
The frequency $\nu_{\rm e, c}$ is usually associated to the peak frequency of the low energy component in
the SED. The injection frequency corresponding to electrons with Lorentz factor $\gamma_{\rm e, min}$ is given by
\begin{align}
    \nu_{\rm e, m} = \frac{4}{3\pi} \frac{q B \delta \gamma_{\rm e, min}^2 }{m_{\rm e} c} \sim 7.5\times10^{15} B_0 \delta_1 \gamma_{\rm e, min, 4}^2 ~~~ {\rm Hz.}
\end{align}
 Another turnover in the synchrotron
spectrum is at the self-absorption frequency $\nu_{\rm SSA}$. Synchrotron self-absorption dominates at
low frequency, specifically in the radio band, and introduces a cut-off like modification around the frequency $\nu_{\rm SSA}$ \citep[ ][]{2014ApJ...789..161N}:
\begin{align}
    \nu_{\rm SSA}^{\prime} \approx \frac{1}{3} \left(\frac{eB^{\prime}}{m_e^3 c} \right)^{1/7} \frac{L_{syn}^{\prime 2/7}}{R^{\prime 4/7}}
    \label{ssa_approx}
\end{align}
where $L_{syn}^{\prime}$ is the synchrotron energy distribution peak luminosity.

The interaction between the photons of the low energy hump and the electrons and positrons producing this hump via synchrotron radiation can produce the second peak in the broadband spectrum (SSC). The peak frequency of this component depends on the the cooling regime of the electrons and on the peak frequency of the synchrotron component. It is given by 
\begin{align}
    \nu_{\rm IC} & = \left \{ \begin{aligned}
        & 2\gamma_{\rm e, min}^2 \nu_{\rm e, m}  ~~~~~~~~~~~~~~~~~~~~&\nu_{\rm e, c} < \nu_{\rm e, m} \\
        & 2\gamma_{\rm e, c}^2 \nu_{\rm e, c} &\nu_{\rm e, m} < \nu_{\rm e, c}
    \end{aligned} \nonumber \right. \\
    &\sim \left \{ \begin{aligned}
        &  1.5\times 10^{24} B \delta_1 \gamma_{\rm e, min, 4}^4 ~ {\rm Hz,}~~&\nu_{\rm e, c} < \nu_{\rm e, m} \\
        &  4.3\times 10^{21} \delta_1 B^{-7} R_{16}^{\prime 4}~ {\rm Hz,}&\nu_{\rm e, m} < \nu_{\rm e, c}
        \label{ic}
    \end{aligned} \right.
\end{align}
for fast and slow cooling respectively. Similarly, the ratio of luminosities of the synchrotron $L_{\rm s}$ and the inverse self-Compton $L_{\rm IC}$ components can be approximated by 
\begin{align}
\frac{L_{\rm SSC}}{L_{\rm syn}} \sim \left \{ \begin{aligned}
    & \frac{2}{3} \tau \gamma_{\rm e, c}^2 \left ( \frac{\gamma_{\rm e, c}}{\gamma_{\rm e,min}} \right )^{1- \alpha_e} & & \nu_{\rm e, c} > \nu_{\rm e, m}\\
    & \frac{2}{3} \tau \gamma_{\rm e, c} \gamma_{\rm e, min} & & \nu_{\rm e, c} < \nu_{\rm e, m}
\end{aligned} \right.
\end{align}
where $\tau = \sigma_T R^{'} n_{\rm e}^{\prime}$ is the opacity of the source for the Compton process and $n_{\rm e}^{\prime}$ is the comoving electron density. It is computed assuming that the Thomson regime is achieved for the peak, which might not always be the case.

For hadronic models (hereinafter HM), and more specifically for proton synchrotron models, the HE component
of the SED is dominated by the proton synchrotron radiation rather than by the inverse Compton scattering.
This model requires that a substantial number of protons are accelerated in the jet to very large Lorentz factor\footnote{In principle, the maximum proton energy
$\gamma_{\rm p}$ could be estimated by assuming an acceleration time of the form
$t_{\rm acc} \sim \gamma_{\rm p} m_{\rm p} c^2 /(\eta c q B)$, where $\eta\sim 1$ is the
acceleration efficiency. This time is then compared to the different cooling time scale to obtain an estimate
of $\gamma_{\rm p, max}$. }. In this case, the required magnetic field is larger than in leptonic models, with $B$ in the order of hundred Gauss. The peak frequency of proton synchrotron emission is at:
\begin{align}
    \nu_{\rm s}^{\rm p} = 4.1 \times 10^{24} B_2 \delta_1 \gamma_{\rm p, max, 9}^2~ {\rm Hz,} \label{eq:proton_sync_mzx_nu}
\end{align}
where we did not consider cooling. In general, hadronic models necessitate much more energetic jets since
they require a large magnetic field, as well as a significant amount of energy in relativistic protons. We further discuss these constraints in Section \ref{disc}.
In addition to synchrotron loses, relativistic protons of the jet also lose energy by photo-pion and Bethe-Heithler photo-pair interactions with the photons.


For hybrid models, a subclass of hadronic models, the low and high energy peaks are explained by leptonic
processes and proton synchrotron emission is required to be subdominant. The requirement on proton
content is obtained by maximizing the neutrino flux at PeV energies, which is constrained by the
radiation from the secondaries produced by the Bethe-Heithler and photo-pion processes.  Indeed, it
has long been speculated that efficient neutrino production is associated with efficient Bethe-Heithler
process, creating a population of HE pairs, which can over-shine the tight constraints in the X-ray band \citep[e.g., order of minutes, ][]{PM15b}.

In order to produce PeV neutrinos, protons should have a comoving energy larger than $E_{\rm p}^{\prime} > 10^{15}/\delta_1$ eV.
Assuming for simplicity that the Bethe-Heithler process creates pairs with Lorentz factor $\gamma_\pm = \gamma_{\rm p}/5$\footnote{This
assumptions requires the inelasticity to be $\kappa_{\rm e} \sim 10^{-4}$. \cite{MPK05} computed the inelasticity and finds that
it steadily decreases from $10^{-3}$ for increasing $\gamma_{\rm p} x$, where $x = h\nu / (m_{\rm e} c^2)$, with $h$ the Planck
constant.}, the pairs created by the protons producing PeV neutrinos are in the fast cooling regime, see Equation \eqref{eq:gamma_c_pm}.
Therefore, the energy produced in the Bethe-Heitler process is efficiently radiated by synchrotron radiation.
For an electron or positron to radiate in X-ray, its Lorentz factor should be 
\begin{align}
    \gamma_\pm^{1\rm keV} = \sqrt{\frac{3\pi \nu c m_{\rm e} }{4B \delta q_{\rm e}}} \sim 5.7\times 10^3 \nu_{1\rm keV}^{\frac{1}{2}} B_2^{-\frac{1}{2}} \delta_1^{-\frac{1}{2}}.
\end{align}
which is smaller than the Lorentz factor of the pairs from the protons producing PeV neutrinos. Therefore, synchrotron radiation from the Bethe-Heitler pairs contributes to the X-ray band. We now estimates the Bethe-Heitler pair spectrum. The Bethe-Heithler pair yield is 
\begin{align}
    \frac{\partial n_\pm}{\partial t} (\gamma_\pm) = 2c \int_0^\infty dx  n_{\rm ph}(x) \int_1^\infty d\gamma_p N_p \frac{d\sigma_\pm}{d\gamma_\pm} \label{eq:pair_yield_CK}.
\end{align}
where $x = h\nu/ (m_{\rm e}c ^2)$ is the photon energy normalised to the electron rest mass. 
Under the head-on approximation and if the photon energy is small enough to neglect proton recoil,
the differential pair rate can be written as \citep{CK13}
\begin{align}
    \frac{d\sigma_\pm}{d\gamma_\pm} \sim \frac{\alpha \sigma_T}{2x \gamma_\pm^2} & & \frac{1}{2x} \leq \gamma_\pm \leq \frac{\gamma_{\rm p}}{2},
\end{align}
where $\alpha$ is the fine structure constant.
We further assume that the photon spectrum is well approximated by
$n_\gamma(\epsilon) = n_{\gamma,0} \epsilon^{-\alpha_{\rm ph}}$, which is realistic since the
synchrotron emission from the electrons forming the low energy bump can be well
approximated by a succession of power-laws with indexes $\alpha_{\rm ph} = 2/3$, $3/2$, $(\alpha_e +1)/2$, where we
neglected self-absorption and specialised to the fast cooling scenario, usually appropriate for HM. We
also further assume that protons do not cool substantially such that their distribution function is 
$N_p = N_{\rm p,0} \gamma_{\rm p}^{-\alpha_{\rm p}}$ for
$\gamma_{\rm p} < \gamma_{\rm p, max}$, then the integral of Equation \ref{eq:pair_yield_CK} yields
\begin{align}
    \frac{\partial n_\pm}{\partial t} (\gamma_\pm) \simeq \alpha c \sigma_T 2^{\alpha_{\rm ph} + 2 -\alpha_{\rm p}} \frac{n_0}{\alpha_{\rm ph}} \frac{N_0}{\alpha_{\rm p}-1} \gamma_\pm^{\alpha_{\rm ph}-\alpha_{\rm p}-1}
\end{align}
Therefore, the pair injection spectrum will be formed of three smoothly connected power-laws with indexes
$\alpha_{\rm ph}-\alpha_{\rm p}-1$,  where $\alpha_{\rm ph} = 2/3$, $3/2$, $(\alpha_e +1)/2$.
Since these pairs are in the fast cooling regime, their distribution function is well
approximated by smoothly connected power-laws with indexes $q= \alpha_{\rm ph}-\alpha_{\rm p}-2$. 
From \cite{RL79}, the resulting photon flux is well approximated by three smoothly connected
power-laws $F_\nu \propto \nu^{-q/2}$.

In proton synchrotron models, when the proton injection index is $\alpha_{\rm p} \sim 2$, 
the specific spectral power $\nu F_\nu$ of the synchrotron emission from the Bethe-Heithler pairs
is nearly flat with indexes $-2/3, -1/4, -(\alpha_{\rm e} -3)/4 $. The spectrum extends up to energies 
\begin{align}
    \nu_{\pm, \rm max} \sim \left (\frac{\gamma_\pm}{\gamma_{\rm p, max}} \right )^2 \frac{m_{\rm p}}{m_{\rm e}}  \nu_{\rm s}^{\rm p} \sim 3.0 \times 10^{26} B_2 \delta_1 \gamma_{\rm p, 9}^2~ {\rm Hz.}
\end{align}
where we used Equation \eqref{eq:proton_sync_mzx_nu} for the synchrotron frequency associated to the highest
energy protons with Lorentz factor $\gamma_{\rm p, max}$.
Yet, because the pair synchrotron emission peaks at such a large frequency, its contribution to the X-ray is
likely to be small and not constraining for proton synchrotron models. However, this is not the case for hybrid
models when the peak frequency for synchrotron radiation from the pairs will be 
\begin{align}
    \nu_{\pm, \rm max} \sim 3.0 \times 10^{19} B_{-1} \delta_1 \gamma_{\rm p, 7}^2~ {\rm Hz,}
\end{align}
around the X-ray frequency, in agreements with the estimates from \cite{PM15b}. It is clear that increasing
the neutrino flux requires to increase the density of protons or of photons. This leads to an increase of
the production rate of pairs, and as a result, a larger synchrotron flux in the X-ray band, which becomes
critical for constraining this type of models \citep{PM15b,GFW19, 2019ApJ...881...46R}.

\section{Modeling of Blazar SEDs} 

\label{sec:applications_TSX}

The code \textit{SOPRANO}, described in Section \ref{sec:intro_soprano}, is used to model the multiwavelength SEDs of
\txs{}, \hsp{} and \source{}. Two of these sources, \txs{} and  \hsp{}, coincide in space and time with the IceCube 170922A and IceCube 200107A events, respectively. The other source, \source{}, shows a prominent flare in the \gray{} band.

It is assumed that protons and electrons are injected in the emitting region with energy distributions
given by Equations \eqref{pdist} and \eqref{edist}, respectively. We also assume that the injection
power-law indexes are such that $\alpha_{\rm e}=\alpha_{\rm p}$. Once injected in the emitting region,
particles interact with the magnetic field and with the photons, producing secondary particles, which themselves
interact, radiate and decay, shaping the broadband SED. The low energy component is interpreted as the
synchrotron emission of the primary electrons while the HE component is formed by joint contributions of
inverse Compton scattering of primary electrons and of synchrotron radiation from the protons, as well as secondary particles
from photo-hadronic interactions. The system of kinetic equations is evolved for one dynamical time scale
$t_{\rm dyn}^{\prime} \sim R^{\prime}/c$ considering the magnetic field to be constant, and taking
into account all relevant processes for particles interactions.

\subsection{Modeling of \txs{} SED}

After the observations of neutrinos from the direction of \txs{} \citep{IceCube18, IceCube18b}, hadronic
processes in its jet have been extensively studied.
The multiwavelength emission and neutrino production were discussed for the $p\gamma$
\citep{2018ApJ...863L..10A,2018ApJ...864...84K, 2018ApJ...865..124M,2019MNRAS.483L..12C,  2019NatAs...3...88G, 2019MNRAS.484.2067R, 2020ApJ...891..115P}
and $pp$ \citep{2018ApJ...866..109S, 2019PhRvD..99f3008L} interaction scenarios.
The current modeling consensus is that the applied one zone models predict, albeit low, but still
consistent results with the observation of one neutrino event in 2017. However, the neutrino flare
in 2014–2015 cannot be explained when both the neutrinos and the electromagnetic emission are produced
from the same region.

\begin{table*}
    \centering
    \caption{Parameter sets used for modeling the SEDs of \txs{}, observed in 2017 and during the neutrino flare in 2014-2015. The electon, proton and magnetic luminosities are also given. } 
   		\begin{tabular}{@{}l c c| c c}
 		& \multicolumn{3}{|c|}{\txs} \\
 		\hline
 		& \multicolumn{2}{|c|}{2017} & \multicolumn{2}{|c|}{2014-2015}\\
 		\hline
		 & Hadronic & Lepto-hadronic & Hadronic & Lepto-hadronic \\
 		\hline
 $\delta$ & $20$  & $20$ & $15$ & $10$ \\
 $R/10^{15}\:{\rm cm}$ & $2.5$ & $10$ & $1$ & $100$  \\
 $B[G]$ & $80$  & $0.57$ & $35$ & $0.65$  \\
 $\gamma_{\rm e, min} $& $100$  & $1000$ & $2\times10^2$ & $9\times10^3$ \\
 $\gamma_{\rm e, cut}$& $2.4\times10^3$  & $4.5\times10^4$ & $10^4$ & =$\gamma_{\rm e, max}$   \\
 $\gamma_{\rm e, max}$& $3\times10^4$  & $6\times10^4$ & $8\times10^4$ & $8\times10^4$\\
 $\alpha_{\rm e}$ & $2.1$  & $2.0$ & $2.0$ & $2.0$ \\
 $\alpha_{\rm p}=\alpha_{\rm e}$ &  2.1 & 2.0 & 2.0 & $2.0$\\
 $\gamma_{\rm p, min}$& 1 & 1 & 1 &1\\
 $\gamma_{\rm p,max}$& $10^9$  & $10^6$ & $2\times10^8$ & $1.2\times10^5$\\\hline
 $L_{\rm e}\:({\rm erg\:s^{-1}})$& $2.2\times10^{44}$  & $9.3\times10^{44}$ & $2.8\times10^{44}$ & $5.3\times10^{44}$\\
 $L_{\rm B}\:({\rm erg\:s^{-1}})$ & $6.0\times10^{46}$  & $4.9\times10^{43}$ & $10^{45}$ & $1.6\times10^{45}$\\ 
 $L_{\rm p}\:({\rm erg\:s^{-1}})$& $2.1\times10^{47}$ & $2.6\times10^{50}$ & $3.4\times10^{47}$ & $4.9\times10^{52}$\\
 		\hline
 		\end{tabular}
 	 \label{tableltxs}
 		\end{table*}

\begin{figure*}
    \centering
    \includegraphics[width=0.45\textwidth]{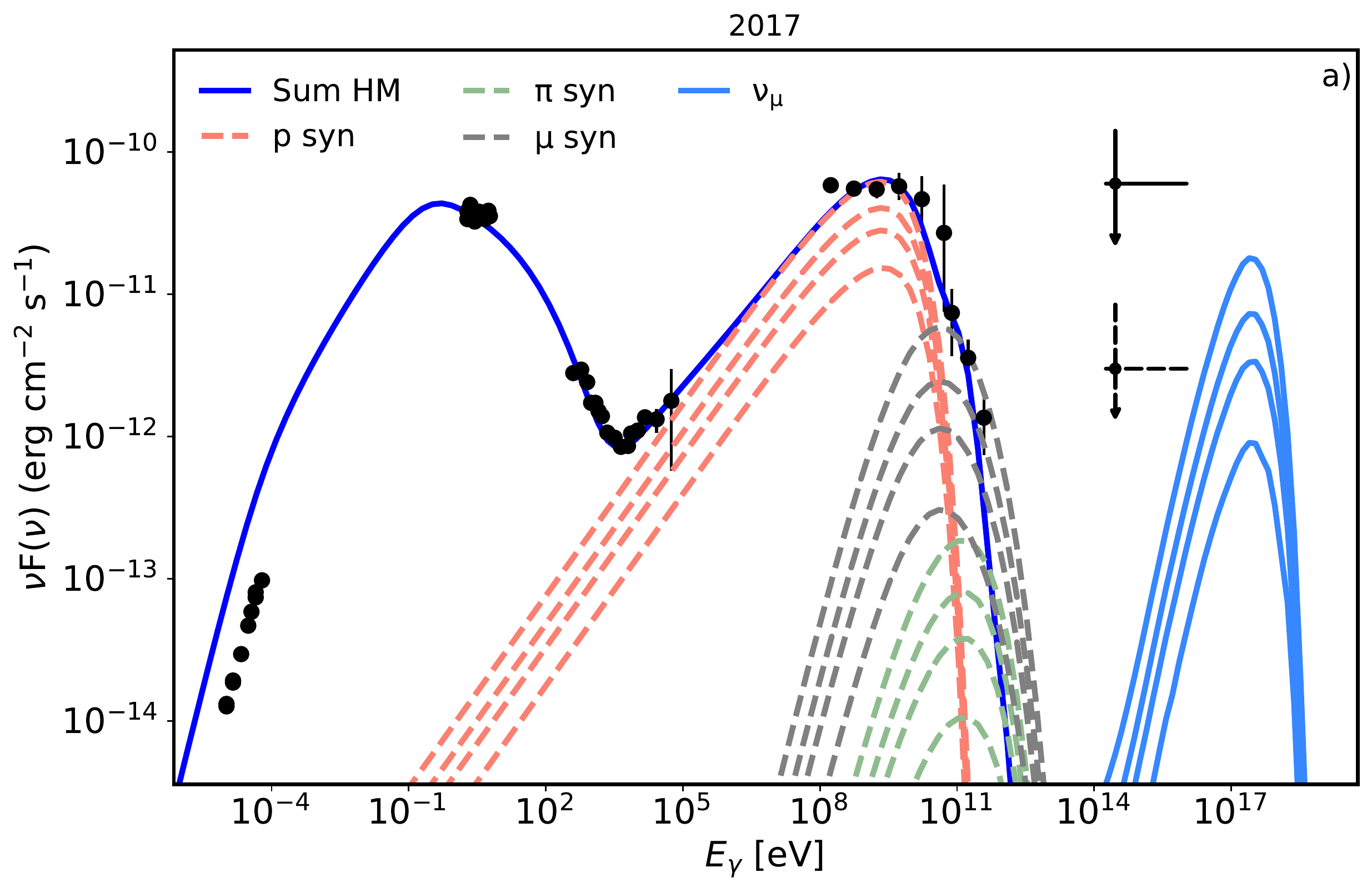} 
    \includegraphics[width=0.45\textwidth]{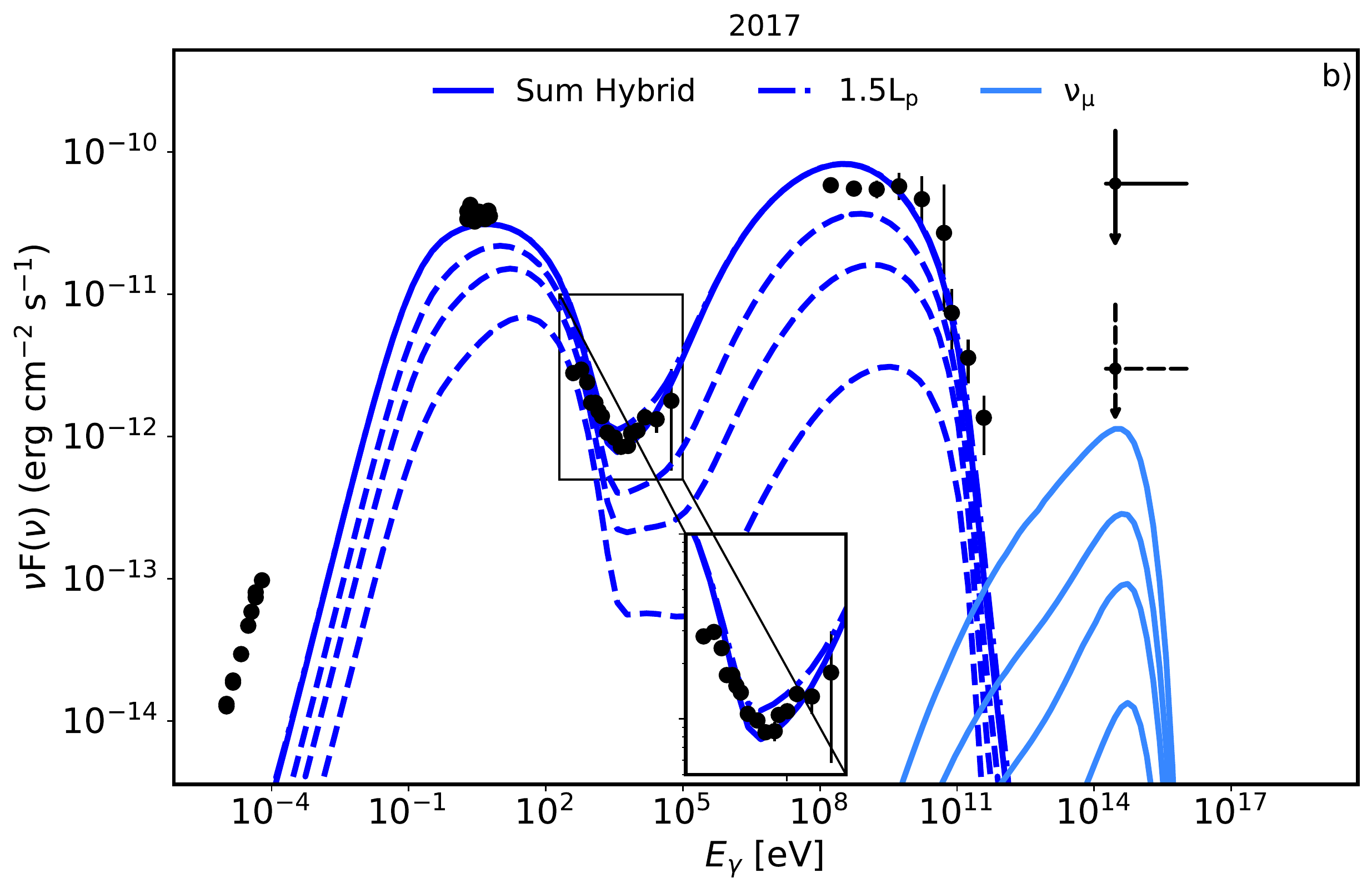}\\
     \includegraphics[width=0.45\textwidth]{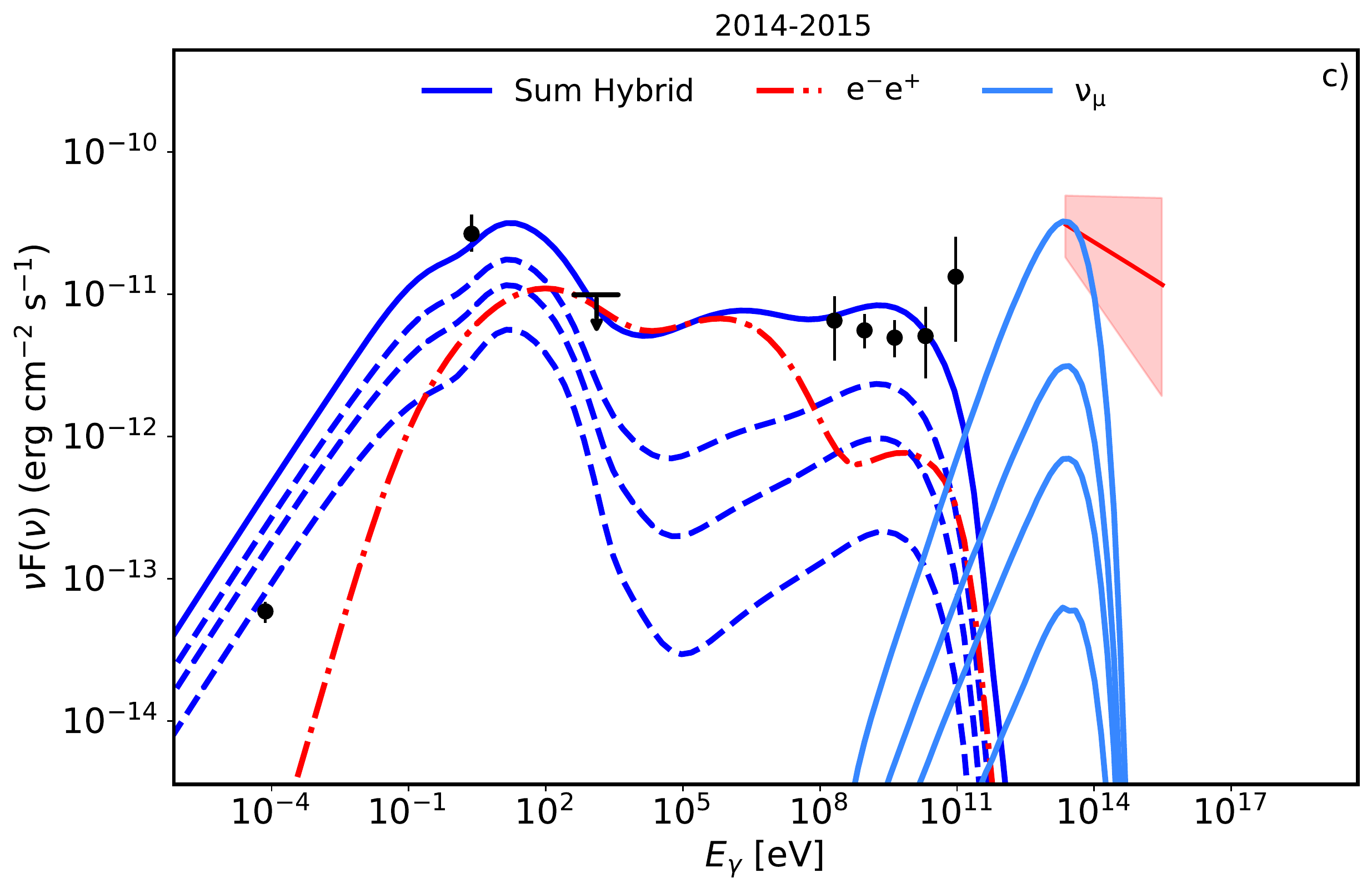}
     \includegraphics[width=0.45\textwidth]{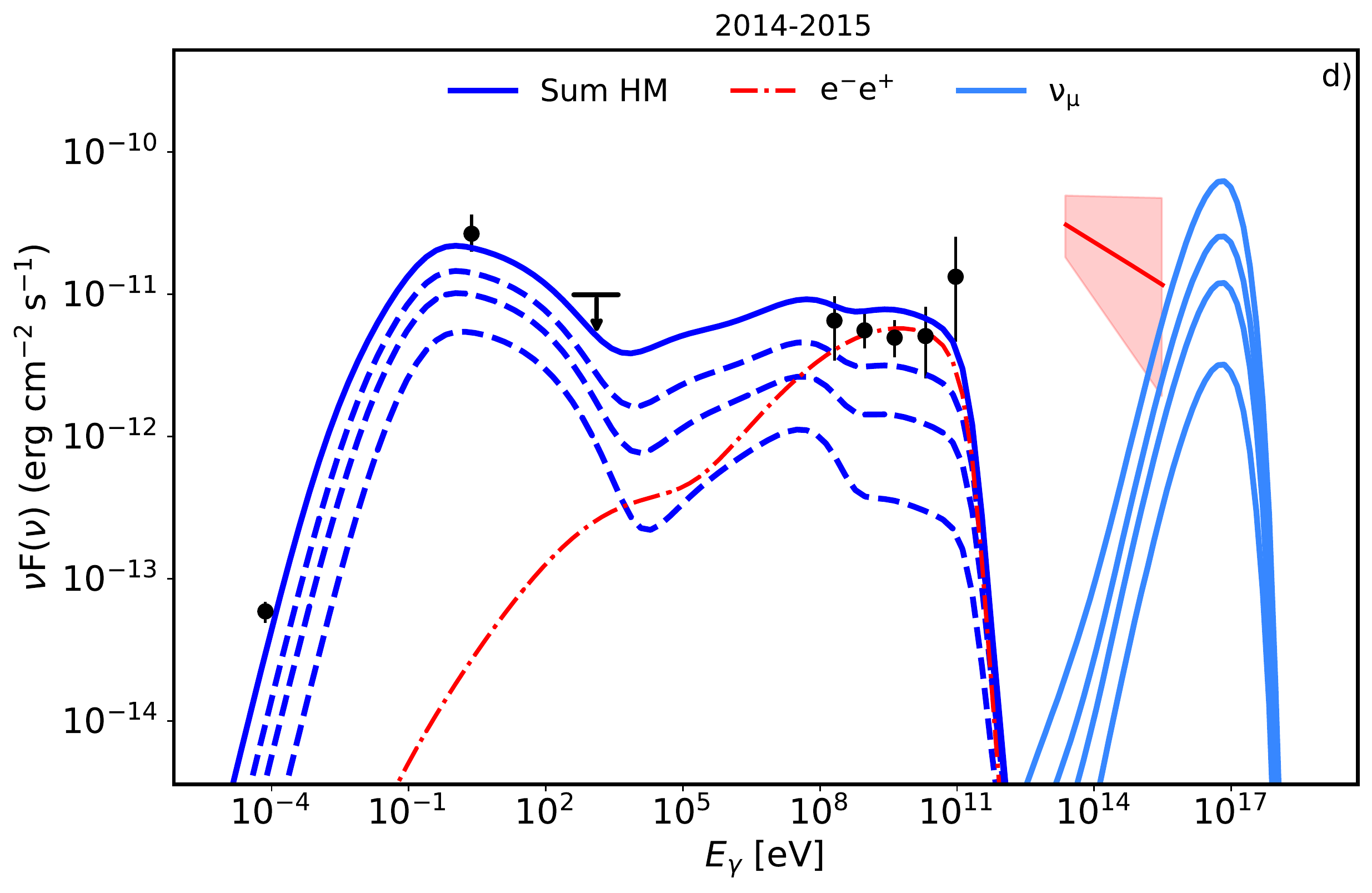}
    \caption{The multiwavelength SED of \txs\ during the neutrino emission in 2017 (upper panels) and during
    the neutrino flare in 2014-2015 (lower panels) modeled within the hadronic and lepto-hadronic hybrid scenarios. The solid blue line in all plots represents the sum of all
    components which has been corrected for EBL absorption considering the model of \citet{2011MNRAS.410.2556D}.}
    \label{txs:SED}
\end{figure*}

Panels a) and b) of Figure \ref{txs:SED} show the SED of \txs\ when the neutrino event was observed.
The multiwavelength data from \citet{IceCube18} are modeled within a HM scenario in panel a).
The corresponding model parameters are given in Table \ref{tableltxs}. The sum of all components, represented by
the blue line in the top left panel of Figure \ref{txs:SED},
satisfactorily explains the observed data. The model over-predicts the radio data, but when taking the synchrotron self-absorption into account via Equation \eqref{ssa_approx}, which is significant below the energies $\approx 10^{-3}$ eV ( $\approx 3\times 10^{11}$ Hz) the model is in agreement with the data.
Under the guise of our modeling, the data up to the soft X-ray band are produced by synchrotron emission of electrons,
which are in the fast-cooling regime. Indeed the magnetic field is required to be high, $B = 80$G, to explain the HE peak with proton synchrotron emission, shown by the red dashed line in panel a) of Figure \ref{txs:SED},
with a contribution of muon synchrotron emission at HEs, represented by the gray dashed line.
The contribution of pion synchrotron emission is negligible and does not contribute
substantially to the flux observed by the MAGIC telescopes \citep{2018ApJ...863L..10A}. The emission in
the transition region between the low and high energy components, in the X-ray band, is dominated by
proton synchrotron emission, with little contribution from the cascade emission of the secondary pairs
produced from the absorption of VHE \grays{} and by the emission of pairs from the Bethe-Heithler process.

The modeling parameters given in the first column of Table \ref{txs:SED} are in the range of similar
estimations for blazars in general and for \txs{} in particular. A Doppler factor $\delta=20$ and a radius 
$R^{\prime}=2.5\times10^{15}$ cm were used in our modeling. This is in agreement with the limits on the variability time of $10^5$s presented in \cite{KMP18} and in \cite{2019MNRAS.484L.104P}. We note that when $\delta=10$ or $15$, the data can also be well reproduced by the model. 
The radius, which defines the density of interacting particles and photons, is a crucial quantity in defining
the type of model. The initial injection power-law index of the emitting electrons is $\alpha_{\rm e} = 2.1$,
a value that can be formed by shock accelerations, \textit{e.g.} \cite{1987PhR...154....1B}. Due to the
high magnetic field, $B = 80$ G, electrons are in the fast cooling regime and their distribution function is a power-law with index $\alpha_{e}+1$. The
initial electron distribution extends up to $\gamma_{\rm cut}=2.4\times10^3$ ($\sim1$ GeV)
which is representative of the acceleration and cooling time scales. Instead, protons cool less efficiently and they
could be accelerated up to much higher energies, \textit{i.e.} $\gamma_{\rm max}=10^9$ ($9.4\times10^{17}$ eV), see
the discussion in Section \ref{disc}.

Previous modelings of \txs{} have shown that hybrid models can be good alternatives to proton synchrotron or
leptonic models \citep{2019MNRAS.483L..12C,GFW19}. They are found to be favourable from the point of view of
neutrino observations. The SED of \txs{}, now modeled within the framework of a hybrid lepto-hadronic scenario,
is shown in panel b) of Figure \ref{txs:SED}.
The model parameters are given in the second column of Table \ref{tableltxs}. The blue dashed
lines represent the time evolution of the spectrum in selected numerical steps, which builds and forms the
overall SED, represented by the solid blue line after one dynamical time scale. 

The synchrotron component peaking between 1-10 eV is up-scattered by the relativistic electrons
to produce the HE and VHE component. The magnetic field in the emitting region is $B=0.57$G
significantly lower than for the proton synchrotron modeling. Therefore, electrons with Lorentz
factor $\gamma_{\rm e, min} = 10^3$ are not substantially cooled in one dynamical time scale.
The electron distribution function is a broken power-law with an exponential cut-off,
$\gamma^{ -\alpha_{\rm e}}_{e}$ and $\gamma^{ -\alpha_{\rm e}+1}_{e} \exp[-\gamma_{\rm e}/\gamma_{\rm e, cut}]$, with
a break at Lorentz factor $\gamma_{\rm e,c} = 7 \times 10^3 $, where we used Equation \eqref{eq:gamma_c_pm}.
In order for the neutrino spectrum to peak around the energy of the observed neutrino
($290$ TeV), the comoving proton distribution function should extend at least up to
$\gamma_{\rm p,max}=10^6 \delta_1^{-1}$. 
This Lorentz factor is lower than what is usually used
in pure HM models. For this hybrid model, protons do not directly
contribute to the observed SED. Their radiative signature is due to the emission of the secondaries of photo-pion
and photo-pair interactions. Their contribution dominate in the X-ray band, which constrains
the proton luminosity and as a consequence the neutrino luminosity. For example, if one increases by 1.5 times theproton luminosity, the model would overshoot the X-ray data, as shown by the dotted-dashed blue line in  panel b) of Figure \ref{txs:SED}.\\
\indent We now present our results of the SED modeling obtained during the historical neutrino flare of \txs{}.
Unfortunately, when 13 $\pm$ 5 neutrinos were observed between October 2014
and March 2015 \citep{IceCube18b}, the multiwavelength coverage is scarce. Yet, the flux upper limit
of $ F < 9.12\times10^{-12}\:{\rm erg\:cm^{-2}\:s^{-1}}$ derived from Swift BAT observations
\citep{2019ApJ...881...46R} introduces substantial difficulties for a one-zone modeling. Indeed, the predicted
number of neutrino events cannot be matched to the IceCube observations.
\citet{2019ApJ...881...46R} and \citet{2019ApJ...874L..29R} have shown that only few neutrino events could
be detected under different optimistic considerations for the emitting region and for the target photon field (internal
or external to the jet). Matching together the observed multiwavelength data and the neutrino data seems to require
two zone models with more free parameters \citep{2019ApJ...881...46R, 2019ApJ...874L..29R}.\\
\indent To accommodate the X-ray limit and try to account for the neutrino flux during this flare, two different assumptions on
the proton distribution function are made. On the one hand, radiation from the secondaries can be constrained to be dominant
in the MeV band, in which there are no observational constraint. On the other hand, radiation from the secondaries could be
dominant in the GeV band and produce the second HE hump. The SEDs of these two models are respectively shown in panels
c) and d) of Figure \ref{txs:SED}, with data from \citet{2019ApJ...874L..29R}. Those two models lead to two very different sets
of parameters for the emitting region, see column 3 and 4 of Table \ref{tableltxs}. The first model requires a large radius
$R^{\prime}=10^{17}{\rm cm}$ and a slowly moving jet with Doppler factor $\delta=10$, while the second model necessitates those parameters
to be $R^{\prime}=10^{15}{\rm cm}$ and $\delta=15$. The required magnetic field also significantly differs between these two models
with $B=35$G for the first model, to be compared to $B=0.65$G for the second one. The first model tends to reproduce the
neutrino flux, albeit produces the peak at lower energies. The second model puts the neutrino peak at larger energy, but
is not able to reproduce the observed neutrino number. In both interpretations, it is clear that the upper limit in the X-ray band imposes strong constraints on the photon spectrum, which in turn limits the proton content in the jet.
Considering larger proton luminosity would lead to over-estimate both the observed \gray{} flux and the X-ray upper limit. 
\subsection{Modeling of \hsp{} SED}

The blazar \hsp{} is another interesting source to study within a hadronic scenario. Indeed, it is a nearby blazar at
redshift $z=0.55703$ \citep{2020MNRAS.495L.108P}, and it lies in the error region of the neutrino event
IC 200107A \citep{2020ATel13394....1G}. The multiwavelength campaign, which started after the neutrino
detection in January 2020, showed that \hsp\ was in a bright X-ray emission state with a synchrotron peak
frequency of $5\times10^{17}$ Hz \citep{2020A&A...640L...4G}. This is a typical value for extreme peak
blazars \citep{2001A&A...371..512C}. It is the first time that the jet of an extreme blazar is associated
with a neutrino event, straightening the assumption that the jets of this blazar type
are potential sites for cosmic rays and even ultra-high energy cosmic ray acceleration
\citep{2016MNRAS.457.3582P}. The multimessenger emission from \hsp{} was interpreted
within various leptonic and lepto-hadronic models by
\cite{2020ApJ...899..113P} and \cite{2020ApJ...902...29P}. \citet{2020ApJ...899..113P}
showed that a change of the X-ray flux above 1 keV does not significantly affect the neutrino flux.
The expected number of neutrinos during the 44-day period is  $6\times10^{-4}$ with a low probability of $\sim0.06$ \% to
detect one or more neutrinos. Alternatively, \citet{2020ApJ...902...29P} investigated the effects of the external photon fields to enhance the neutrino production.

\begin{table*}
    \centering
    \caption{Parameters used to model the multiwavelength SEDs of \hsp{} and \source{}. The electron, proton and magnetic luminosity is also displayed.}
   		\begin{tabular}{@{}l c c c| c c}
 		&  \multicolumn{4}{|c|}{\hsp{}} & \multicolumn{1}{|c|}{\source{}} \\
 		\hline
 		& \multicolumn{2}{|c|}{January 8th} & \multicolumn{2}{|c|}{January 10th}\\
 		\hline
		 & Hadronic & Lepto-hadronic & Hadronic & Lepto-hadronic & Hadronic \\
 		\hline
 $\delta$ & 15 & 30 & 15 & 30 & 55 \\
 $R/10^{15}\:{\rm cm}$ & 0.3 & 10 & 0.3 & 10 & 0.32 \\
 $B[G]$ & 45 & 0.11 & 45 & 0.08 & 70 \\
 $\gamma_{\rm e, min} $ & $10^4$ & 100 &  $5\times10^3$ & $100$ & 1 \\
 $\gamma_{\rm e, cut}$& $6\times10^5$ & $2\times10^6$  & $2\times 10^5$ & $7\times 10^5$ & $2.4\times 10^2$\\
 $\gamma_{\rm e, max}$& $9\times10^5$ & $6\times10^6$ & $5\times10^5$ & $6\times10^6$ & $4\times10^2$\\
 $\alpha_{\rm e}$ & 1.9 & 2.0 & 1.9 & 2 & 1.8 \\
 $\alpha_{\rm p}=\alpha_{\rm e}$ & 1.9 & $2.0$ & 1.9 & 2 & $1.8$\\
 $\gamma_{\rm p, min}$& 1 &1 & 1 & 1 & 1\\
 $\gamma_{\rm p,max}$& $9\times10^8$ & $10^6$ & $9\times10^8$ & $10^6$ & $2.1\times10^8$\\\hline
 $L_{\rm e}\:({\rm erg\:s^{-1}})$ & $1.2\times10^{44}$ & $1.6\times10^{44}$ & $7.3\times10^{43}$ & $2.1\times10^{44}$ & $1.9\times10^{44}$\\
 $L_{\rm B}\:({\rm erg\:s^{-1}})$ & $1.5\times10^{44}$ & $4.1\times10^{42}$ & $1.5\times10^{44}$ & $2.2\times10^{42}$ & $5.7\times10^{45}$\\ 
 $L_{\rm p}\:({\rm erg\:s^{-1}})$& $3.2\times10^{46}$ & $8.0\times10^{50}$ & $3.2\times10^{46}$ & $1.8\times10^{51}$ & $1.3\times 10^{49}$\\
 		\hline
 		\end{tabular}
 	 \label{tablel3HSP}
 		\end{table*}


\begin{figure*}
    \centering
    \includegraphics[width=0.48\textwidth]{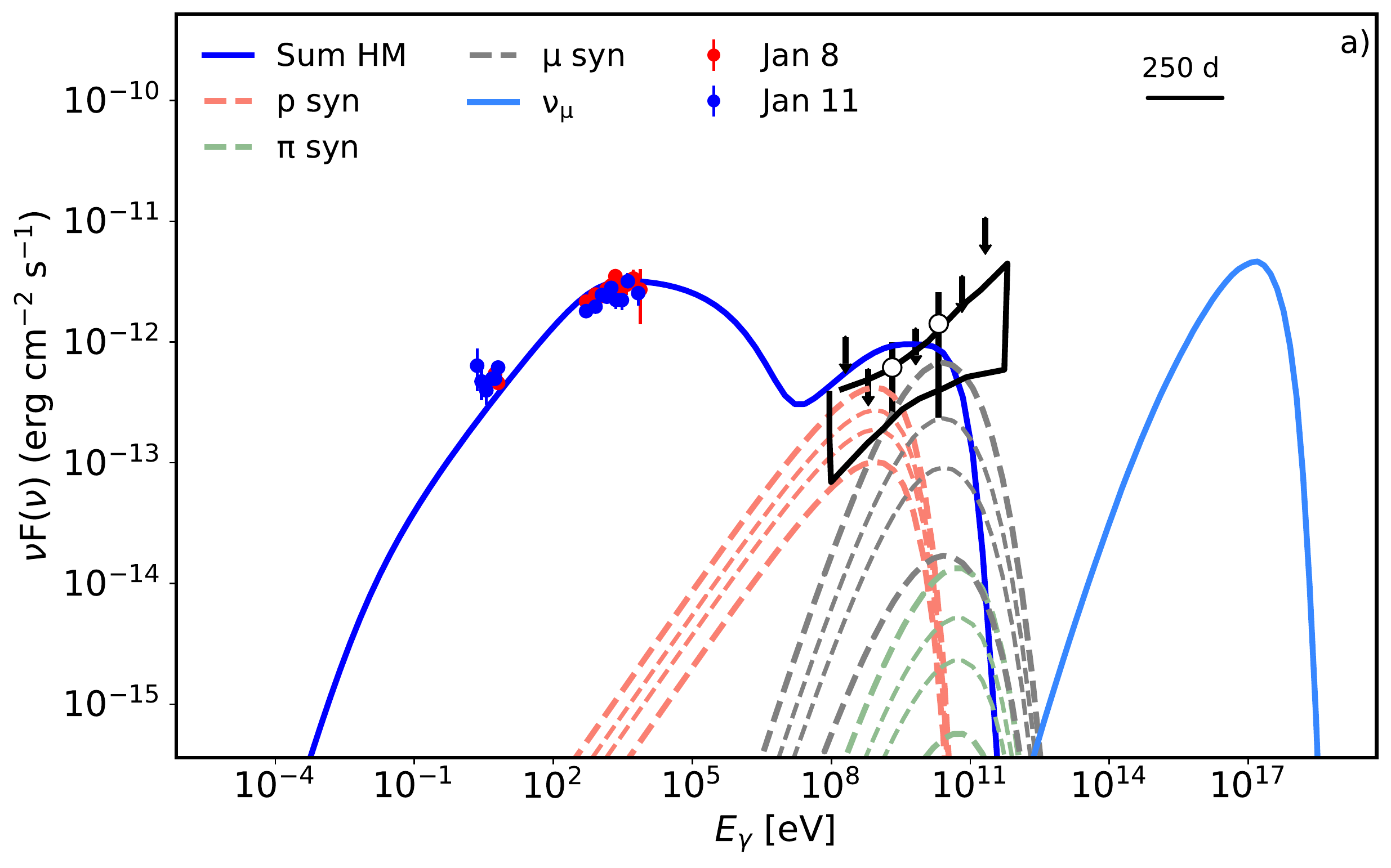} 
    \includegraphics[width=0.48\textwidth]{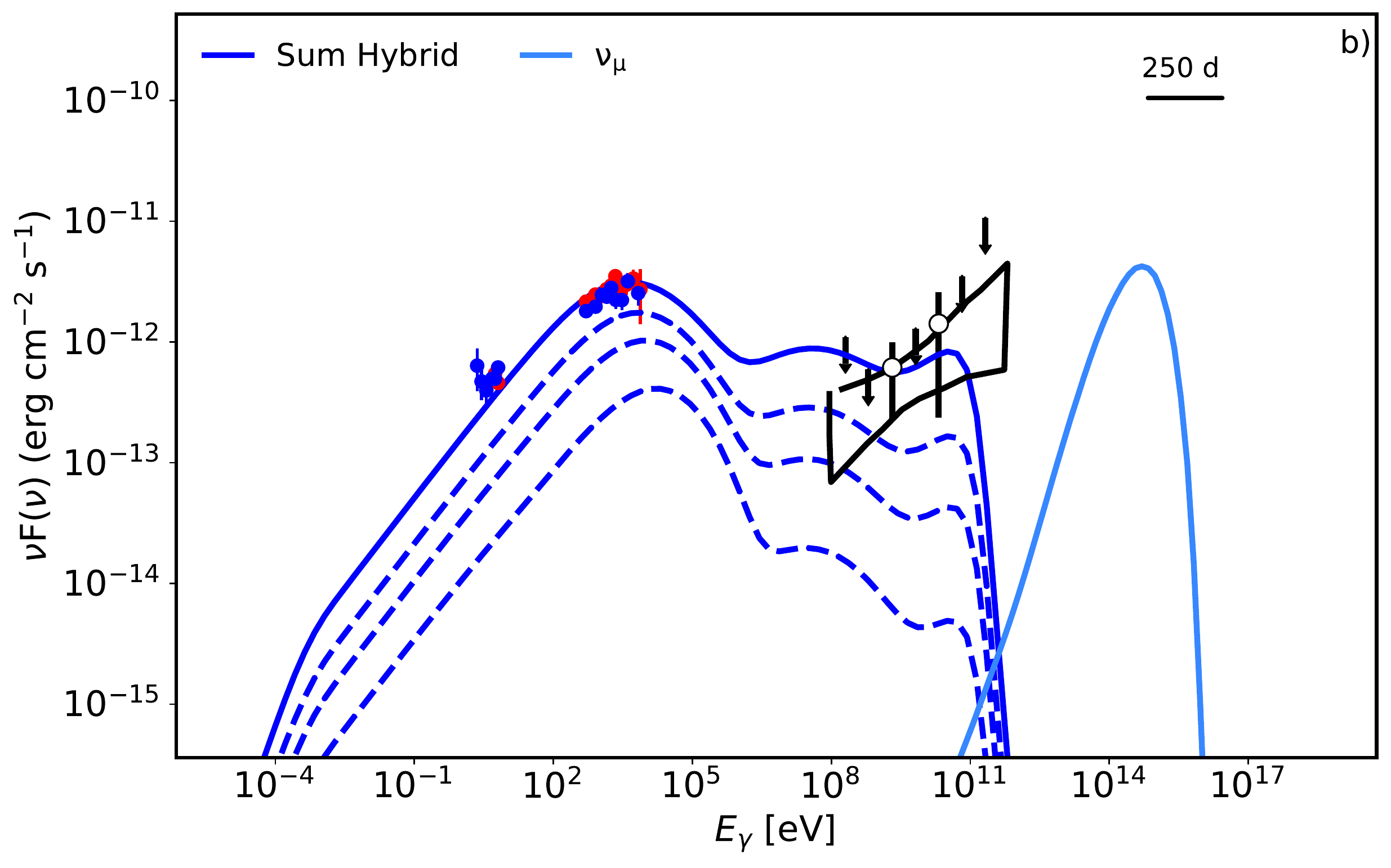}\\
    \includegraphics[width=0.48\textwidth]{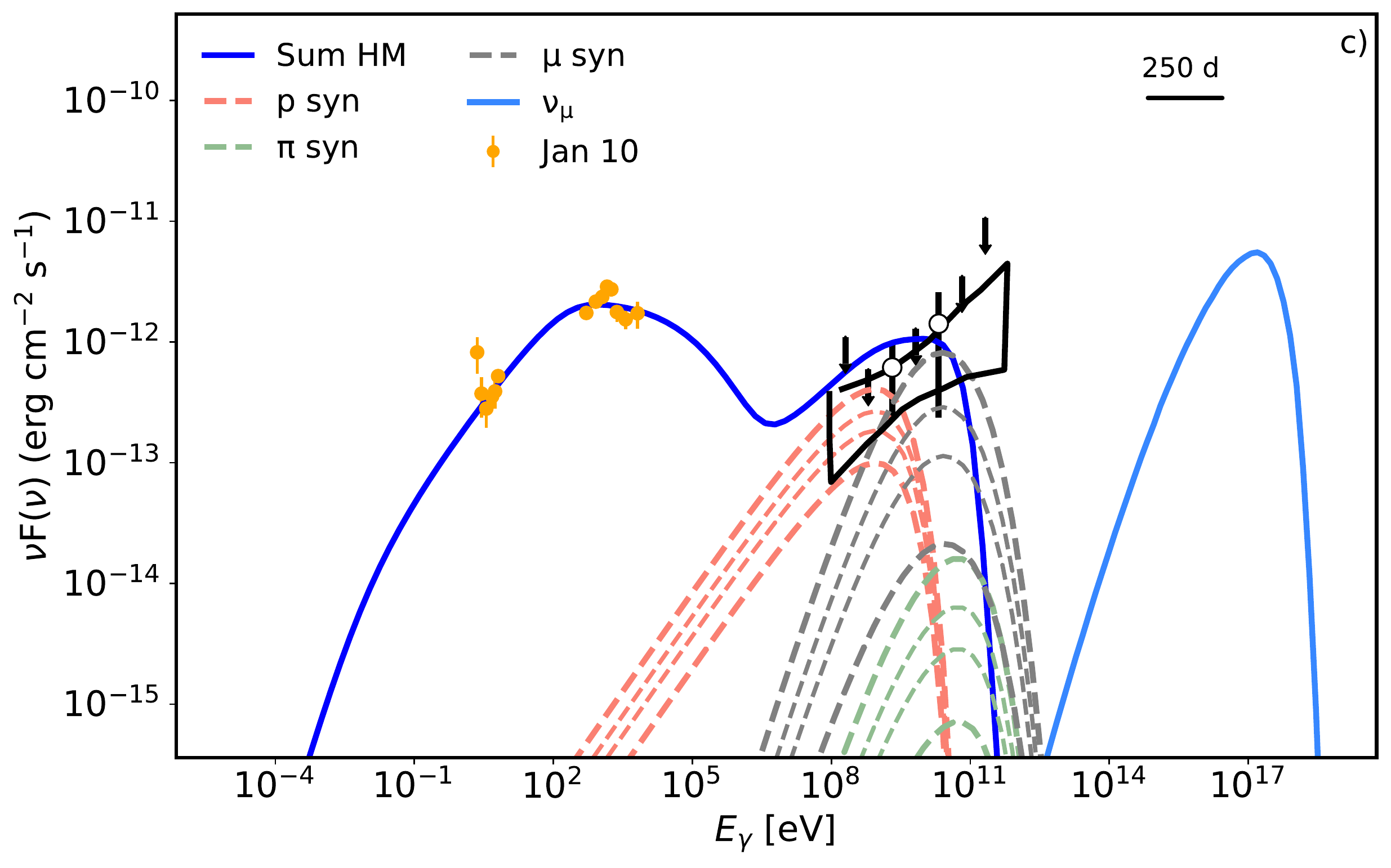}
    \includegraphics[width=0.48\textwidth]{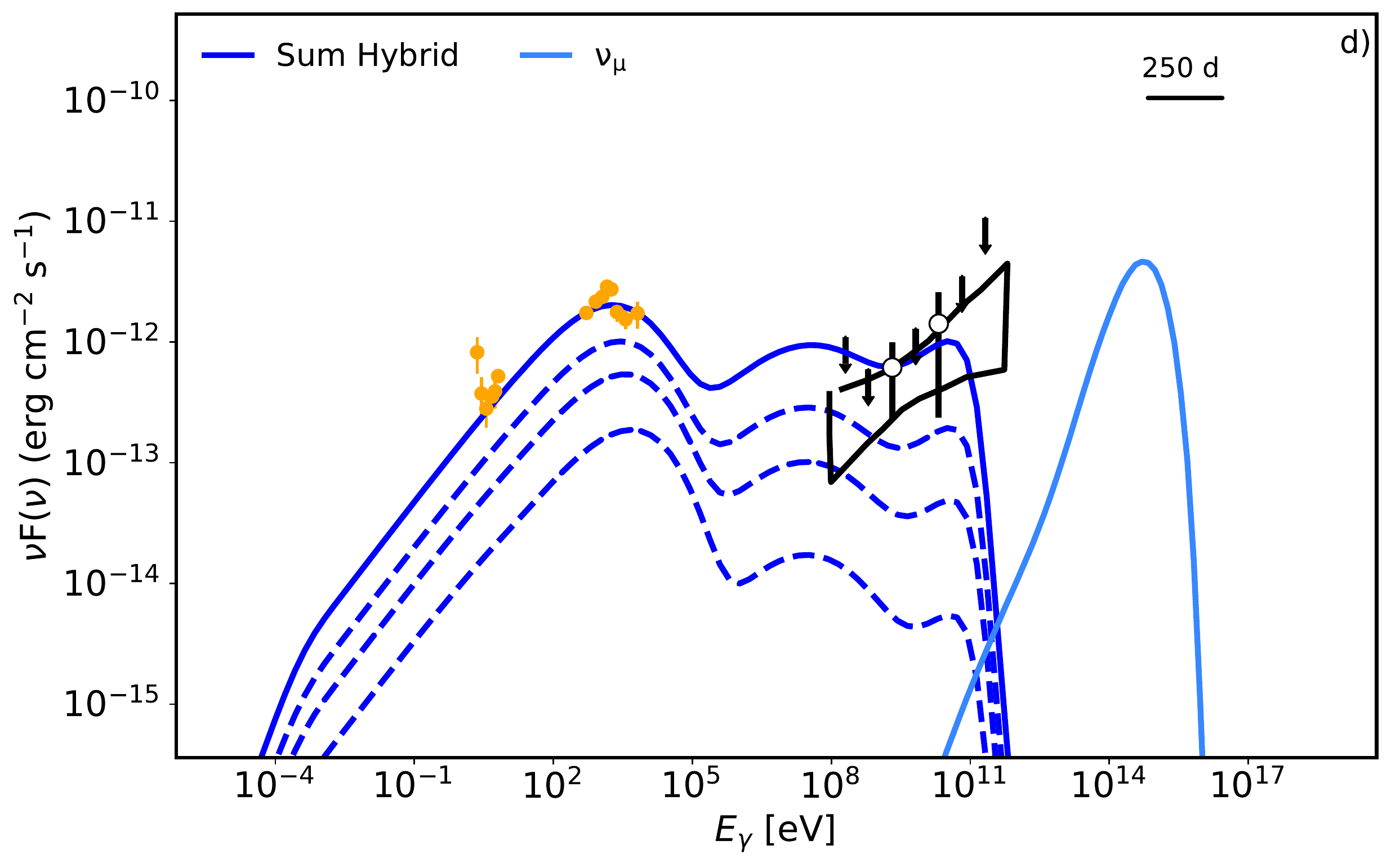}
    \caption{The multiwavelength SEDs of \hsp{} on the the 8$^{\rm th}$
    and the 11$^{\rm th}$ of January (panels a) and b) on the top raw) and on the 10$^{\rm th}$ of January
    (panels c) and d) on the bottom raw). The data are taken from
    \citet{2020A&A...640L...4G,2020ApJ...899..113P}. The observed spectrum including all processes
    is shown by the solid blue line. All models (solid blue lines) have been corrected for EBL
    absorption considering the model of \citet{2011MNRAS.410.2556D}.}
    \label{sed:hsp}
\end{figure*}
The SED of \hsp{} is shown in Figure \ref{sed:hsp}, where the multiwavelength data are from 
\citet{2020A&A...640L...4G}. Optical, UV and X-ray data were acquired on the 8$^{\rm th}$, 10$^{\rm th}$
and 11$^{\rm th}$ of January. However, since the data taken on the 8$^{\rm th}$ and 11$^{\rm th}$ of January
seem to have the same flux and spectral shape \citep{2020ApJ...899..113P}, we only model the data from the
8$^{\rm th}$. The lack of available multiwavelength data does not allow to constrain the low and high energy
peaks, which hardens the estimation of the model free parameters. A hint of a $20-30$ minutes variability has been
found in the NICER and NuSTAR data, but only at the $\sim3.5\:\sigma$ level \citep{2020ApJ...902...29P}.
Therefore, the compactness of the emitting region cannot be constrained. \citet{2020ApJ...899..113P}
investigated the blob radius–Doppler factor relation for a wide range of photo-pion production efficiency and
for several set of parameters. In order to keep the generality,
in the current study, the SED of \hsp{} is modeled for two different parameter configurations. For the HM, we consider $R^{\prime}=3\times10^{14}$ cm and $\delta=15$, while for the lepto-hadronic model we assume 
$R^{\prime}\simeq10^{16}$ cm and $\delta=30$.

In our hadronic modeling, the HE component is mainly due to the synchrotron emission of protons, shown
by the red dashed line in panel a) of Figure \ref{sed:hsp}. Protons are assumed to have an energy distribution
$N_{\rm p} \propto \gamma_{\rm p}^{ - 1.9}$ and to be accelerated up to $\gamma_{\rm p,max}=9\times10^8$,
corresponding to $8.4\times10^{17}$ eV. At VHEs, the largest contribution is due to muon synchrotron radiation,
represented by the gray dashed line in panel a) of Figure \ref{sed:hsp}. 
The high synchrotron peak at $\sim10^4$ eV can be reproduced when
$\gamma_{\rm e, cut}=6\times10^5$ and $B=45$ G.
The minimal energy of the accelerated electrons
is relatively high, $\gamma_{\rm e, min} = 10^4$, but still in the range of parameters usually estimated for
ultra-high-frequency-peaked blazars, see \textit{e.g.} \citet{2015MNRAS.448..910C}. For the hybrid lepto-hadronic
modeling, shown in panel b) of Figure \ref{sed:hsp}, the emitting electrons should be
accelerated up to $\gamma_{\rm e, cut} = 2\times10^6$ so the SSC component extends to the GeV band
to explain the observed data. In this model,
a lower magnetic field of $0.11$ G is required because of the larger radius of the emitting region ($R^{\prime} = 10^{16}$ cm).
The emission of the secondary pairs from protons accelerated up to $\gamma_{\rm p,max}=10^6$ dominates in
the sub-MeV band.

The hadronic and hybrid modeling of the SED observed on the 10$^{\rm th}$ of January 2020,
is displayed in panel c) and d) of Figure \ref{sed:hsp}, respectively. 
Since the peak
of the low energy component, defined by the X-ray data, is at lower energies than for observations performed on the 8$^{\rm th}$ of January, the modeling requires a three times smaller cutoff energy, \textit{i.e.}
$\gamma_{\rm e, cut} = 2\times10^5$ and $\gamma_{\rm e, cut} = 7\times10^5$ for the
hadronic and lepto-hadronic modelings, respectively. The other parameters are given in Table \ref{tablel3HSP} and are similar with these obtained from modeling the data observed on the 8$^{\rm th}$ of January.

\subsection{Modeling of \source{} SED during the 2015 flare}

\begin{figure}
    \includegraphics[width=0.45\textwidth]{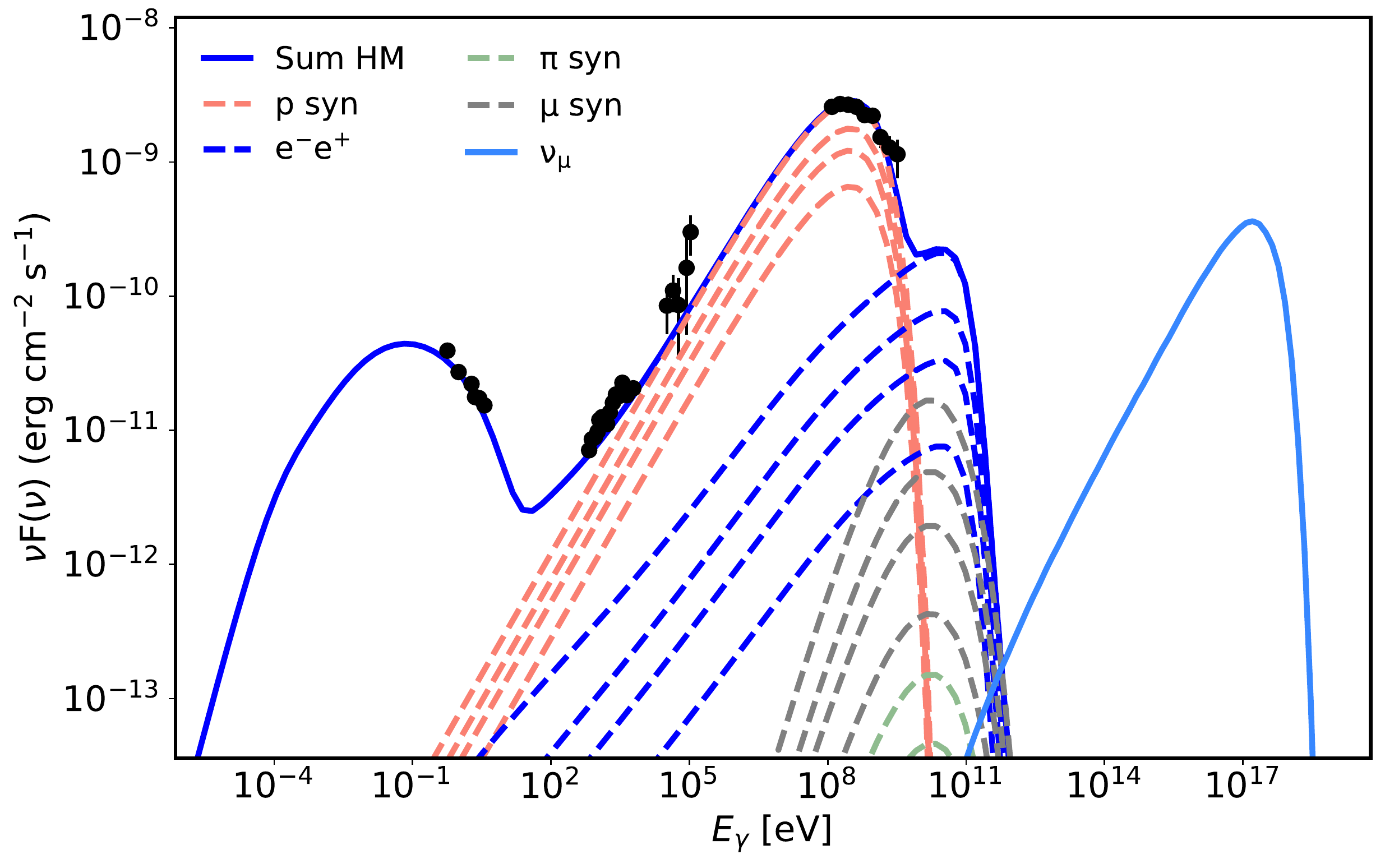}
    \caption{The multiwavelength SED of \source{} during the exceptional flaring activity in 2015. The contribution of different particle emission is shown by dashed lines whereas the thick solid line represents the observed spectrum, corrected for EBL absorption considering the model of \citet{2011MNRAS.410.2556D}. }
    \label{c:sed}
\end{figure}

The emission from the powerful FSRQs \source\ at redshift $z=0.536$ has been detected in all
possible spectral bands. Its broadband emission is characterized by high amplitude
variability almost in all energy bands \citep[e.g., order of minutes, ][]{2014A&A...567A..41A} and in particular in the HE \gray\ band, which present the fastest variability. 
On the 16$^{\rm th}$ of June 2015, Fermi LAT observations showed that \source\ was in an exceptionally bright state.
The flux increased up to $3.6\times10^{-5}\:{\rm photon \: cm^{-2}\:s^{-1}}$ with a flux doubling time on the
order of 5 minutes \citep{2016ApJ...824L..20A}. IceCube performed a time-dependent neutrino signal search
correlated with this \gray{} flare but no evidence for a signal was found
\citep{2021ApJ...911...67A}.
We use \textit{SOPRANO} to model the SED of \source{} during its flare to infer the neutrino flux. We consider
a HM and explain the second peak with proton synchrotron emission. The parameters of our modeling are given
in Table \ref{tablel3HSP}.


Figure \ref{c:sed} shows the multiwavelength SED of \source{} taken from \citet{2016ApJ...832...17B},
alongside with the results of our modeling. During the brightening, the X-ray emission of the source appears
with a hard photon index $<1.50$,
smoothly connecting with the INTEGRAL data, described by a power-law with index  $1.08$ \citep{2016ApJ...832...17B}. In the
HE \gray\ band, the spectrum presents a  power-law with photon index $2.21$ with a turn over 
\citep{2015ApJ...808L..48P}. We make the hypothesis that the HE component
is produced from a single mechanism. In our modeling it is interpreted as proton synchrotron emission, represented by the
red dashed lines in Figure \ref{c:sed}. This interpretation requires that protons are accelerated up
to $\gamma_{\rm p, max} = 2.1 \times 10^8$, see Table \ref{tablel3HSP}. 
The compactness of the emitting region implies a high efficiency for photo-pion and photo-pair interactions, which inject
energetic secondary pairs. The contribution of these pairs dominates above $\sim10$ GeV and peak at
$\sim 100$ GeV, as can be seen  by the blue dashed line in Figure \ref{c:sed}. The synchrotron radiation of
the primary electrons peaks at $\sim 0.1$ eV and its HE tail accounts for the observed
optical/UV data. These data constrain the cut-off energy to be relatively low, $\gamma_{\rm e, cut}=2.4\times10^2$, otherwise,
for $B = 70$ G and $\delta = 55$, the synchrotron radiation
would overshoot the observed flux in optical and UV bands. A similar hadronic modeling for this flare
is presented in \cite{2016ApJ...832...17B} and in \cite{2017MNRAS.467L..16P}.

\section{Discussion}
\label{disc}
The primary class of objects to be studied in the multimessenger context are blazars which are
associated with neutrino events observed by IceCube. Even if the associations are not at
the 5$\sigma$ significance level, the observations by IceCube put
some constraints on the physical processes taking place in relativistic jets. Using the hadronic
time-dependent model constrained by their neutrino emission,
the broadband SEDs of two blazars, namely \txs{} and \hsp{}, respectively associated to the neutrino
events IC 170922A and IC 200107A, are studied. We also analysed the SED of \source\ during its 2015 \gray\
flaring period. For each source, we present several modelings assuming that different components dominate
in the HE \gray\ band. For the sources studied in this paper, 
we find that the proton synchrotron model, the secondaries
emission model and the hybrid lepto-hadronic model  
can explain the observed SEDs under sensible assumptions for the particle energy distributions. 

Some of our modelings require a compact emitting region, with $R^{\prime} \lesssim 10^{15}$cm. In principle, the maximum energy
of the particles, and specifically of the protons, is limited by requiring their Larmor radius, given by
$r_{\rm p, L} = \gamma_{\rm p} m_{\rm p} c^2/(q B)$,
to be smaller than the emission region. All our models are consistent with this requirement, and therefore protons
can be accelerated to the maximum energy as given in Tables \ref{txs:SED} and \ref{tablel3HSP}. The strongest constraints are obtained for
the hadronic model of \hsp{} for which the protons with $\gamma_{\rm p, max} = 9 \times 10^8$ have Larmor radius $r_L = 6.3 \times10^{13}$cm,
while the emitting region has size $R = 3\times 10^{14}$cm. The maximum particle energy can also be
limited by synchrotron cooling. Specifically, \cite{dHM96} balanced acceleration time-scales for shock and
gyroresonant acceleration with cooling time scale via the synchrotron process to find that the electrons
can be accelerated up to 
\begin{align}
   \gamma_{\rm e}^{\rm max} \sim 4 \times 10^7 B^{-\frac{1}{2}}.
\end{align}
For all our modelings, we have $\gamma_{\rm e, max} < \gamma_{\rm e}^{\rm max}$. Only the lepto-hadronic
modeling of \hsp{} is marginally consistent with this limit. Such constraints are highly dependent on the
acceleration mechanism and vary for alternative scenarios, such as particle acceleration by magnetic reconnection or 
in shear layers.


The power-law index of the accelerated particles, assumed to be equal for protons and electrons, is found to be in the range $\alpha=1.8-2.1$, a value in agreement with prediction of shock acceleration theories \citep[e.g., order of minutes, ][]{KGG00,2012ApJ...745...63S}. This index is defined by the acceleration processes, and we note that protons
and electrons could have different indexes. In fact, if particles are accelerated by shocks, the properties of their acceleration depend on the direction of the shock with the magnetic field: quasi-parallel shocks accelerate
both ions and electrons,  while quasi-perpendicular shocks only accelerate electrons
\citep{CS14a, CS14b, CS14c,PCS15, GSN14}. It is worth noting that magnetic reconnection could be the
mechanism accelerating particles in blazar jets, see \textit{e.g.} \citet{GU19}, in which case it is
also expected that the power-law of accelerated electrons and protons be slightly different \citep[e.g., order of minutes, ][]{GLL16}). 

An important quantity allowing to compare and contrast the models is the luminosity
carried out by electrons, protons and  the magnetic field. They are respectively computed
with Equations \eqref{eq:Lb_jet}, \eqref{eq:Lp_jet} and \eqref{eq:Le_jet} and
are given in Table \ref{txs:SED} for \txs{} and Table \ref{tablel3HSP} for \hsp{} and \source{}.
In all our models, the total luminosity of the jet is defined by
the proton content. This is expected since our models are designed to produce
a high neutrino flux. Specifically, in the case of \txs{}, for the proton synchrotron
model, shown on panel a) of Figure \ref{txs:SED}, the required
luminosity for the jet is $L_{\rm tot}=L_{\rm p}+L_{\rm e}+L_{\rm B}=2.7\times10^{47}\:{\rm erg\:s^{-1}}$.
The energy budget in the emitting region is dominated by the particles $(L_{\rm p}+L_{e})/L_{\rm B}\simeq3.5$,
yet the system is closed to equipartition. For this model, the required luminosity exceeds by one order of magnitude the Eddington luminosity
of $\simeq4\times10^{46}{\rm erg\:s^{-1}}$ for a black hole mass of $3\times10^8\: M_\odot$, as estimated for
\txs{} using the absolute R-band magnitude \citep{2019MNRAS.484L.104P}. This is in agreement with previous
studies suggesting
that in the case of proton synchrotron models or models producing a high neutrino flux, the required jet
luminosity exceeds that of the Eddington limit \citep{2019ApJ...871...81X}.
Within a lepto-hadronic model, matching the neutrino flux with the neutrino event of \txs{}, displayed on panel b) of Figure \ref{txs:SED}, requires the jet luminosity to be $\sim10^{50}\:{{\rm erg}\:{\rm s}^{-1}}$, significantly exceeding that of the Eddington luminosity.
Although the Eddington luminosity is  not a strict limit and super-Eddington luminosities were previously reported \citep{2019ApJ...880...67J}, this value is extremely large. On the other hand, for the neutrino flare
in 2014-2015, when assuming that the emission from the secondary pairs solely dominates in the X-ray and
\gray{} bands, an unrealistically high luminosity of $\sim10^{52}\:{\rm erg\:s^{-1}}$ is obtained. Indeed, matching the high neutrino flux with the large radius ($10^{17}$ cm) imposed by the modeling in this case, requires a large protons density, hence the too large proton luminosity. In the alternative interpretation,
when the emission from the secondary pairs dominates in the GeV band, a modest
luminosity of $3.4\times10^{47}\:{\rm erg\:s^{-1}}$ is estimated.

For \hsp{}, the situation is identical to that of \txs{}. For the HM,
a luminosity of $3.2\times10^{46}\:{\rm erg\:s^{-1}}$ is estimated while the lepto-hadronic modeling
requires the jet luminosity to be $\sim10^{51}\:{\rm erg\:s^{-1}}$. The black hole mass of \hsp{} was estimated,
using two different methods, to be $3\times10^8\: M_\odot$ \citep{2020MNRAS.495L.108P} or $\sim8\times10^8\: M_\odot$
\citep{2020ApJ...902...29P}. Therefore, the luminosity estimated from the hadronic modeling is compatible with
the Eddington luminosity $(4-10)\times10^{46}\:$erg$\:$s$^{-1}$. In principle, the proton contribution
to the overall jet luminosity can be decreased by assuming that protons have a softer energy distribution,
$\alpha_{\rm p}>2.0$, different than that of the electrons. However, this introduces a new free parameter for the modeling, and the difference for the proton luminosity would only be a factor of a few.

For \source{}, the hadronic interpretation of the SED observed in 2015 is natural,
considering the difficulties encountered by the leptonic models. Indeed, when considering
external inverse Compton scenario, the interpretation  of the observed large Compton dominance
($\sim70$, the luminosity ratio of the high- and low -energy components) requires a strongly matter
dominated jet \citep{2015ApJ...808L..18A}. In the alternative hadronic modeling, the data from the X-ray
band to the \gray{} band can be well reproduced by proton synchrotron emission, provided they are efficiently accelerated
up to energy $2\times10^{17}$ eV with a power-law index $-1.8$. The modeling requires a relatively high
jet total luminosity $\sim10^{49}\:{\rm erg\:s^{-1}}$, which exceeds the Eddington luminosity ($\sim10^{47}\:{\rm erg\:s^{-1}}$) for a black hole mass of $8\times10^8\: M_\odot$ \citep{2009A&A...505..601N}.
However, this required luminosity is not a strong argument to disfavour the hadronic origin of \source{} emission during the 2015 flare considering that the source was in an exceptionally bright state.

Having estimated the model parameters of each blazar SEDs, the corresponding neutrino
flux can be derived. The flux of muon neutrino, $F_{\nu_{\mu}}(E_{\nu_{\mu}})$,
in all considered scenarios is shown by the light blue line in Figures \ref{txs:SED}-\ref{c:sed}. When available,
the neutrino flux is compared with the limit imposed by the IceCube detector. This flux can be transformed to the
expected observed number of neutrinos in the IceCube detector using its averaged effective area
$A_{\rm eff}({\rm E_{\mu}})$, which is mostly a function of the incident neutrino energy. For \hsp{}
and \source{}, the average area from \citet{2019EPJC...79..234A}
was cosidered, while for \txs{} we used the area released after the observation of IceCube-170922A
\footnote{https://icecube.wisc.edu/science/data-releases/}. The effective area increases with energy and
reaches its maximal value for energies above several hundreds of PeV. The expected number of muon neutrinos and anti-neutrinos is computed through 
\begin{equation}
        N_{\nu_{\mu}+\bar{\nu}_{\mu}}=t_{\rm exp}\:\int_{E_{\rm min, \nu_{\mu}}}^{E_{\rm max, \nu_{\mu}}}\:F_{\nu_{\mu}}(E_{\nu_{\mu}})\:A_{\rm eff}({\rm E_{\nu_\mu}})\:dE_{\nu_{\mu}}
\end{equation}
where the minimum and maximum energy of the neutrinos are $E_{\rm min, \nu_{\mu}}=100$ GeV and
$E_{\rm max, \nu_{\mu}}=10^9$ GeV, respectively, chosen to correspond the limits for the effective area.
The expected number of neutrino events depends on the duration of the source activity, $t_{\rm exp}$, over which
the neutrinos are emitted. The neutrino oscillation, within the quasi-two neutrino oscillation assumption, is taken into account by 
\begin{equation}
    N_{\nu_{\mu}}^{obs} = 0.575 N_{\nu_{\mu}} + 0.425 N_{\nu_{e}},
\end{equation}
where $N_{\nu_{\mu}}^{obs}$ is the observable distribution of muon neutrinos, while, $N_{\nu_{\mu}}$
and $N_{\nu_{e}}$ are the emitted muon and electron neutrino distributions
\citep{2018arXiv180205781F}.

The expected number of neutrinos during the 6 months flare of \txs{} is 0.43 and 0.23 for the hadronic and
the lepto-hadronic scenarios, respectively. During the 2014-2015 neutrino flare, our most
optimistic model predict 3.0-3.3 neutrinos for a 6 months exposure time (note however that the IceCube
observational window was $\sim110$ days). However, this lepto-hadronic modeling requires the jet luminosity to reach unrealistic values, $4.9\times10^{52}\:{\rm erg\:s^{-1}}$, significantly exceeding the Eddington limit. 
By slightly varying the model parameters, a higher neutrino event count can be estimated,
but it always remains below the $13 \pm 5$ events mark. The neutrino flux directly depends on the proton
content in the jet, which is limited by the upper limit of the X-ray luminosity. Our results are in
agreement with previous estimations for \txs{} and confirm that within a one-zone scenarios, $13 \pm 5$
events from the direction of \txs{} cannot be explained \citep{2019ApJ...881...46R, 2019ApJ...874L..29R}.

In the case of \hsp{}, the muon neutrino rate,
$N_{\nu_{\mu}+\bar{\nu}_{\mu}}/t_{\rm exp}$, is within $6\times10^{-4}-4.8\times10^{-3}$ per day. This implies that
under this rate of emission, the expected number of neutrinos to be detected by IceCube in a time corresponding
to the duration of the flare is very low. By exploring different parameter sets,
\citet{2020ApJ...899..113P} concluded that in the most promising scenarios, there is a $\sim1$\% to $\sim3$\% to
observe one neutrino over the time characteristic of the long-term emission of \hsp{} (years).
As the neutrino emission seems to coincide with extreme behaviour of \hsp{} in the X-ray band,
in principle, a large number of neutrinos could be expected if such an activity continues for a longer period. 
However, this is not the case for \hsp{}. Similarly, when considering the flaring activity of \source{}, a
neutrino daily rate as high as $0.15$ per day is estimated.
However, for a relatively short period of the source activity, from minutes to one day, no neutrino events in
the IceCube detector are expected, in agreement with \cite{2021ApJ...911...67A}. 

\section{Conclusion}\label{conc}
Extensive multiwavelength data campaigns from radio to TeV energy bands and simultaneous observations of VHE neutrinos by increasingly more precise experiments pave the way towards a better understanding of highly energetic sources, both in terms of emission mechanism and dynamics. The understanding of the broad SED and neutrino emission requires detailed time-dependent numerical models of the interactions between leptons, photons and hadrons. We have presented a new kinetic model of photo-hadronic and leptonic interactions aiming at studying the emission of optically thin (for Compton scattering) scenarios of relativistic sources (e.g., AGNs and GRBs). Our numerical solution of the kinetic equations for protons, neutrons, pions, muons, neutrinos, pairs, and photons conserves the total energy of the system as well as the number of particles where required. The code takes as an input the spectral injection rate of the particles (e.g., electrons and/or protons), and compute the time evolution of all relevant particles, including the secondaries, as they interact and cool, allowing the computation of the broadband emission spectrum at any given period.

In this paper, we have applied \textit{SOPRANO} to model the SEDs of three blazars, two of which are potentially associated to neutrino emission observed by IceCube. We have assumed different models for the production of the HE component 
and compute in all cases the expected number of muon neutrinos. The time-dependent nature of the code allowed to follow the evolution of all particles in one dynamical time scale and then assess the proton content in the jet by using the radiative spectrum of either secondaries or initial particles. This is necessary for the estimation of the expected number of neutrinos. Such time-dependent treatment of the particle evolution enabled us to constrain different scenarios of neutrino production by using the limits imposed by the observations in different bands.

\section{acknowledgements}
NS and SG acknowledge the supported by the Science Committee of RA, in the frames of the research project No 20TTCG-1C015. SG acknowledges the hospitality of the Max Planck Institute for Physics in Munich, where part of this research was completed and the support from German Academic Exchange Service (DAAD) in Armenia via short-term scholarship. DB was supported by the Deutsche Forschungsgemeinschaft (SFB 1258) when most of the work was done, and presently acknowledges support from the European Research Council via the ERC consolidating grant $\sharp$773062 (acronym O.M.J.).
\section*{Data availability}
The data underlying this article will be shared on reasonable request to the corresponding author.
\bibliographystyle{mnras}
\bibliography{biblio}


\onecolumn
\appendix

\numberwithin{equation}{subsection}

\section{Physical processes in \textit{Soprano} and their kinetic equations}

\label{sec:kinetic_equation}


In the current version of \textit{SOPRANO}, the isotropic kinetic equations for photons, electrons and positrons (considered as one species, see below), protons, neutrons, charged and neutral pions, muons, neutrino and anti-neutrino of all relevant\footnote{$\tau$ neutrino cannot be produced by photo-hadronic interactions.} flavors are evolved in time. For the photon distribution function, we assign $n_{\rm ph}$ to be the number of photons per unit volume per hertz. We further define $N_{\rm i}$ to be the number of particles of i species per unit volume per unit Lorentz factor of particle i. Here, i can be all leptons and all hadrons. Finally, we define $N_{\nu_{\rm i}}$ as the number of neutrinos of i flavour per unit volume per GeV. In our numerical approach, all hadrons and leptons are considered relativistic with $\gamma_{\rm i} \geq 1$. This appendix gives an overview of the kinetic equations, of the cross-sections and of the kinetic equations used in \textit{SOPRANO} for all considered interactions. In Appendix \ref{sec:numerical_discretization},
we detail the numerical prescription.


\subsection{Kinetic equations for all particles}
Here, we summarize all terms appearing in the kinetic equations for all particle species.
We denote $Q$, $S$ and $C$ as the source, sink and cooling terms, respectively. The contribution of inverse Compton scattering is denoted by $R_{\rm IC}$ for the photons and it is a cooling term for the leptons. Detailed expressions for the
interaction kernels are given in the next subsections of this appendix.
\begin{itemize}
    \item \textit{Photons} are produced by the synchrotron radiation of all charged particles and by the
    decay of neutral pions, $\pi_0$. They are absorbed by pair production and redistributed in energy
    by inverse Compton scattering. We neglect the absorption of photons in the photo-pion and photo-pair processes.
    We did not consider synchrotron self-absorption and are planning to include it in the next version. The
    resulting kinetic equation takes the form
    \begin{equation}
        \frac{\partial n_{\rm ph}}{ \partial t} = - S_{\gamma\gamma \rightarrow e^{+}e^{-}} + Q_{\pi_0} + R_{\rm IC} + \sum_{i \in [p, \mu^\pm, \pi^\pm, e^\pm]} Q_{\rm synch}^{i} ,
    \end{equation}
    where the last sum runs on all charged particles. 
    \item \textit{Leptons} (electrons and positrons) are considered as a single species. They are created by muon decay,
    Bethe-Heitler photo-pair production and two photons recombination. They also undergo synchrotron
    cooling such that the final kinetic equation reads as
    \begin{equation}
        \frac{\partial N_{\rm e^{ \pm}}}{ \partial t} =  Q_{\mu^\pm} + Q_{p\gamma \rightarrow e^+ e^-} + Q_{\gamma\gamma \rightarrow e^{+}e^{-}} + C_{\rm IC} + C_{\rm synch} .
    \end{equation}
    \item \textit{Protons} are loosing energy by synchrotron emission, photo-pair and
    photo-pion interactions. Protons are produced through photo-hadronic interactions between photons and neutrons,
    and are turned to neutrons for a substantial fraction of photo-pion interactions. The kinetic equation
    takes the form 
    \begin{equation}
        \frac{\partial N_p}{\partial t} = C_{p\gamma \rightarrow p \pi} + C_{p\gamma \rightarrow e^+ e^-} + C_{\rm synch} - S_{\gamma p \rightarrow n \pi } + Q_{\gamma n \rightarrow p \pi} .
    \label{peq}
    \end{equation}
    \item \textit{Neutrons} are produced in photo-pion interactions and turned to protons by the same process.
    The kinetic equation takes the form 
    \begin{equation}
        \frac{\partial N_n}{\partial t} = - S_{n\gamma \rightarrow p\pi} + Q_{p\gamma \rightarrow n \pi } + C_{n\gamma \rightarrow n \pi} .
    \end{equation}
    In the current version of the code, we do not include neutron decay. Indeed, for the very large particle Lorentz factor
    involved, neutrons would escape the source before decaying. In principle, neutrinos produced by neutron
    decay should contribute to the observed overall signal. But since we are considering models in which the
    neutron number is always much smaller than the proton number, we can safely neglect this contribution. Note that
    numerically investigating a model similar to that of \cite{AD03} would require a proper treatment of neutron decay.
    \item \textit{Charged pions}, $\pi_+$ and $\pi_-$, are produced by photo-pion interactions. Then, they cool
    via synchrotron emission and decay. The kinetic equation for both species takes the form
    \begin{equation}
        \frac{\partial N_{\pi_\pm}}{\partial t} = Q_{p\gamma \rightarrow \pi} + Q_{n\gamma \rightarrow \pi} - S_{\pi} + C_{\rm synch}.
    \end{equation}
    The kinetic equations were solved independently for $\pi^+$ and $\pi^-$ since the branching ratio in
    photo-pion production is different for negative and positive pions. This impacts the
    production ratio between the different neutrino species further.
    \item \textit{Neutral pions} have a kinetic equation similar to that of charged pions but without
    synchrotron cooling.
    \item \textit{Muons} are produced from the decay of charged pions. They lose energy by synchrotron radiation and
    decay. Therefore, the kinetic equation is
    \begin{equation}
        \frac{\partial N_{\mu_\pm}}{\partial t} = Q_{\pi_\pm}  - S_{\mu_\pm} + C_{\rm synch}.
    \end{equation}
    \item \textit{Muon and electron neutrinos and anti-neutrinos} are produced in the decay of pions and muons. We consider the two flavours independently, but neutrino and anti-neutrinos of the same flavour are combined.
    \begin{align}
            \frac{\partial N_{\nu}}{\partial t} = Q_{\pi_\pm}  + Q_{\mu_\pm} .
    \end{align}
\end{itemize}
For each of the processes, the details of the terms $Q$, $S$, $C$ and $R$ are given in the next subsections of Appendix \ref{sec:kinetic_equation}
together with the cross-sections used in \textit{SOPRANO}.

\subsection{Synchrotron emission and cooling}

In \textit{SOPRANO}, all charged particles lose their energy by synchrotron radiation as soon as a magnetic field is specified. Our current treatment does not include synchotron self-absorption, which will be added
in a future update. For each charged particles, we describe the evolution of the distribution function due 
to synchrotron loses by a diffusion equation in energy space
\begin{equation}
    \frac{\partial N_{\rm i}}{\partial t} = \frac{1}{m_{\rm i} c^2} \frac{\partial }{\partial \gamma_{\rm i}} \left (  N_{\rm i} \int_0^\infty j_{\rm synch} (\nu,\gamma_{\rm i}) d\nu \right ) , \label{eq:synch_kinetic_equation_electron}
\end{equation}
while the photon kinetic equation is given by an integro-differential type equation:
\begin{equation}
    \frac{\partial n_{\rm ph}}{\partial t} = \int_1^\infty N_{\rm i}(\gamma_{\rm i}) \frac{j_{\rm synch}}{h\nu} (\nu, \gamma_i) d\gamma_{\rm i}.  \label{eq:synch_kinetic_equation_photon}
\end{equation}
The synchrotron emissivity $j_{\rm synch}$ is given in the relativistic approximation by 
\begin{equation}
    j_{\rm synch} (\nu) = \frac{\sqrt{3}q^3 B}{ m_{\rm i} c^2}  \int_0^{\pi/2} \sin(\theta_p) F(X) d\theta_p  
\end{equation}
with $X = \nu/\nu_c$, 
\begin{equation}
    \nu_c = \frac{3}{4 \pi} \gamma_{\rm i}^2  \frac{q B}{m_{\rm i} c}   \sin(\theta_p),   
\end{equation}
and
\begin{equation}
    F(X) = X\int_X^{\infty} K_{5/3} (\xi) d\xi
\end{equation}
with $K_{5/3}$ the modified Bessel function. This expression fails when the particle Lorentz factor tends towards one,
in which case expression suitable with cyclo-synchrotron should be used \citep{MNY96,MM03}. Therefore, in our
numerical models, synchrotron emission due to mildly-relativistic particles is inaccurate. In practice,
this parameter space is not relevant for blazars or for optically thin emission models of GRBs.

\subsection{Inverse Compton scattering.}

For the rate of Compton scattering of an electron with Lorentz factor $\gamma$ interacting with an isotropic distribution of photons of energy $x_1 = h \nu_1 / (m_e c^2)$, we consider the relativistic approximation given by \cite{Jon68}
\begin{align}
    &R \left (\gamma, x_1\rightarrow x_2  \right ) \equiv \frac{dN}{dt dx_2} = \frac{3 c}{4} \frac{\sigma_T }{x_1 \gamma^2} \left [ 2 q \ln(q) + (1+2q)(1-q) + \frac{1}{2} \frac{(4x_1 \gamma q)^2}{1+4x_1 \gamma q} (1-q)  \right ] ,
\end{align}
where $x_2 = h \nu_2 / (m_e c^2)$ is the energy of the scattered photons, and
\begin{equation}
    q = \frac{x_2}{4x_1 \gamma^2 \left ( 1 - \frac{x_2}{\gamma} \right ) } ,
\end{equation}
is limited to $q< 1$ and  $q > 1/(4\gamma^2)$. This approximation to the exact cross-section is often used for
blazar modeling. It it is accurate for large electron Lorentz factors, relevant for those objects. This
approximation also implies that electrons can only lose energy and photons can only gain energy. Therefore,
it is not suitable to describe the heating of electrons by the photon field.

For the relativistic electrons considered in \textit{SOPRANO}, the kinetic equation takes the form of a diffusion equation
\begin{equation}
    \frac{\partial }{\partial t} \left ( N_{\rm e^\pm} \right ) = \frac{1}{m_e c^2} \frac{\partial }{\partial \gamma_e}  \left ( P_c N_{e^\pm} \right )  \label{eq:ke_electron_CS},
\end{equation}
where the power radiated by Compton scattering is 
\begin{equation}
    P_c(\gamma) =  m_e c^2 \int_{x_1} \int_{x_2} dx_1 dx_2  R \left (\gamma, x_1\rightarrow x_2  \right ) n_{\rm ph}(x_1) (x_2 - x_1).
\end{equation}
On the other hand, we preserve the full integro-differential expression for the photon kinetic equations
since for each inverse Compton scattering off relativistic electrons, photons gain a large amount of
energy compared to their initial energy : 
\begin{align}
    \frac{\partial n_{\rm ph}}{\partial t} (x_2) =&  \int_\gamma \int_{x_1} d\gamma dx_1 R(\gamma,x_1 \rightarrow x_2)  N_{\rm e^\pm}(\gamma) n_{\rm ph} (x_1) -  n_{\rm ph} (x_2) \int_\gamma \int_{x_1} d\gamma dx_1 R(\gamma,x_2 \rightarrow x_1)  N_{\rm e^\pm}(\gamma) .    \label{eq:kinetic_IC_photon}
\end{align}
The first term represent the redistribution of photons of energy $x_1$ to $x_2$ and the second term represent the redistribution of photons of energy $x_2$ to all other possible energies.

\subsection{Pair production}

For the pairs, the kinetic equation of the photon-photon annihilation process reads
\begin{equation}
    \frac{\partial N_e}{\partial t} = c \int_{x_1} \int_{x_2} n_{\rm ph}(x_1) n_{\rm ph}(x_2) \sigma_{\rm  2\gamma \rightarrow e^\pm}(x_1, x_2 \rightarrow \gamma) dx_1 dx_2.
\end{equation}
For the photons, the kinetic equation can be written
\begin{equation}
    \frac{\partial n_{\rm ph}}{\partial t}(x_1) = - n_\gamma(x_1) \int_{x_2} n_{\rm ph} (x_2) \sigma_{\rm 2\gamma \rightarrow e^\pm}^0 (x_1 , x_2) dx_2 ,
\end{equation}
where 
\begin{equation}
    \sigma_{\rm 2\gamma \rightarrow e^\pm}^0 = 2 \int_\gamma \sigma_{\rm 2\gamma \rightarrow e^\pm} (x_1, x_2, \gamma) d\gamma .
\end{equation}
For the cross-section, we use the formula given by \cite{BS97}. It is recall here for convenience
\begin{equation}
    \sigma_{\rm 2\gamma \rightarrow e^\pm}(x_1,x_2 \rightarrow \gamma) = \frac{3}{4} \frac{\sigma_T c}{x_1^2 x_2^2} \left . \left ( \frac{\sqrt{E^2 - 4 \alpha_{\rm cm}^2}}{4} + H_+ + H_- \right )\right|_{\alpha_{\rm cm}^L}^{\alpha_{\rm cm}^U},
\end{equation}
where the center of mass energy $\alpha_{\rm cm}$ is given by 
\begin{equation}
    \alpha_{\rm cm} = \sqrt{\frac{x_1 x_2}{2}},
\end{equation}
and 
\begin{align}\left\{
\begin{aligned}
    E &= x_1 + x_2 \\
    c_{\pm} &= (x_{1,2}-\gamma)^2 - 1 \\
    d_{\pm} &= x_{1,2}^2 + x_1 x_2 \pm \gamma (x_2 - x_1) \\
    \alpha_{\rm cm}^{a,b} &= \sqrt{1/2} \sqrt{\gamma(E-\gamma) + 1 \pm \sqrt{\left[\gamma (E-\gamma) + 1\right]^2 - E^2}} \\
    \alpha_{\rm cm}^U &= \min \left (\sqrt{x_1 x_2} , \alpha^a_{\rm cm} \right ) \\
    \alpha^L_{\rm cm} &= \max \left (1, \alpha_{\rm cm}^b \right )
    \end{aligned} \right.
\end{align}
Finally for $c\neq 0$, the $H$ functions are defined by 
\begin{align}
    H_{\pm} =&  - \frac{\alpha_{\rm cm}}{8\sqrt{x_1 x_2 + c_{\pm} \alpha_{\rm cm}^2}} \left ( \frac{d_{\pm}}{x_1 x_2} + \frac{2}{c_{\pm}}\right ) +  \frac{1}{4} \left ( 2 - \frac{x_1 x_2 -1}{c_{\pm}} \right ) I_{\pm} + \frac{\sqrt{x_1 x_2 + c_\pm \alpha_{\rm cm}^2}}{4} \left ( \frac{\alpha_{\rm cm}}{c_\pm} + \frac{1}{\alpha_{\rm cm}x_1 x_2}\right ),
\end{align}
where 
\begin{align}
    I_\pm = \left \{ 
    \begin{aligned}
    &\frac{1}{\sqrt{c_\pm}} \ln \left ( \alpha_{\rm cm} \sqrt{c_{\pm}} + \sqrt{x_1 x_2 + c_\pm \alpha_{\rm cm}^2} \right )&  & c_\pm > 0, \\
    &\frac{1}{\sqrt{c_\pm}} \arcsin \left ( \alpha_{\rm cm} \sqrt{-\frac{c_\pm}{x_1x_2}}\right )& & c_\pm < 0,
    \end{aligned}\right.
\end{align}
while for $c_\pm =0$, we have
\begin{align}
    H_\pm =& \left ( \frac{\alpha_{\rm cm}^3}{12} - \frac{\alpha_{\rm cm} d_\pm}{8} \right )\frac{1}{\left ( x_1 x_2\right )^{3/2}} + \left ( \frac{\alpha_{\rm cm}^2}{6} + \frac{\alpha_{\rm cm}}{2} + \frac{1}{4 \alpha_{\rm cm}}    \right ) \frac{1}{\sqrt{x_1 x_2}} .
\end{align}

\subsection{Bethe-Heitler pair production}
Photo-pair production, also called Bethe-Heitler process, is the creation of an electron positron
pair by the interaction between a proton and a photon. The cross section of this process is given
by the formula 3D-2000 of \cite{MOK69}, see also \cite{Blu70} and \cite{KA09}, which we recall
here for convenience:
\begin{align}
\frac{d^2 \sigma }{dE_- d \mu} &=  \left ( \frac{\alpha Z^2 r_0^2 p_- p_+}{2k^3}  \right ) \left [-4\sin^2(\theta)\frac{2E_-^2+1}{p_-^2 \Delta_-^4} +\frac{5E_-^2 - 2E_+ E_- +3}{p_-^2 \Delta_-^2} + \frac{p_-^2 - k^2}{T^2 \Delta_-^2} +\frac{2E_+}{p_-^2 \Delta_-} \right. \nonumber \\
+ & \frac{Y}{p_- p_+} \left ( 2E_- \sin^2(\theta) \frac{3k+p_-^2E_+}{\Delta_-^4} + \frac{2E_-^2(E_-^2+E_+^2)-(7E_-^2+3E_+E_- +E_+^2)+1}{\Delta_-^2}  + \frac{k(E_-^2 - E_- E_+ -1)}{\Delta_-}  \right ) \label{eq:BH_total_cross_section_gould}\\
- & \left. \frac{\delta_+}{p_+ T} \left( \frac{2}{\Delta_-^2} -\frac{3k}{\Delta_-} - \frac{k(p_-^2 - k^2)}{T^2 \Delta_-}    \right ) - \frac{2y_+}{\Delta_-}   \right ], \nonumber
\end{align}
where 
\begin{align}
E_+ & = k - E_- & p_+ &= \sqrt{p_+^2-1}  & p_- & = \sqrt{E_-^2-1}  \\
T &= \sqrt{k^2+p_-^2-2kp_- \cos(\theta)} & & & Y& = \frac{2}{p_-^2} \ln \left ( \frac{E_+E_- + p_+p_- +1}{k}  \right ) \\
y_+& = \frac{1}{p_+} \ln \left ( \frac{E_+ + p_+}{E_+ - p_+} \right )  &&& \delta_+ &=\ln \left( \frac{T+p_+}{T-p_+} \right ).
\end{align}

The kinetic equation for the production of pairs is given by
\begin{align}
    \frac{\partial }{\partial t} \left ( N_{\rm e^\pm} (\gamma_{\rm e}) \right ) &= c \int_{\gamma_{\rm p} m_{\rm p} > \gamma_{\rm e} m_{\rm e}}  dE_{\rm p} N_{\rm p} \frac{dN_{\rm e}}{dE_{\rm e}} .
\end{align}
The pair spectrum is given by \cite{KA09}
\begin{align}
    \frac{dN_e^-}{dE_e} &= \frac{1}{2\gamma_p^3} \int_{x =\frac{(\gamma_p + E_e)^2}{4\gamma_p^2 E_e}}^\infty \int_{\omega = \frac{(\gamma_p + E_e)^2}{2\gamma_p E_e}}^{2\gamma_p x} \times  \int_{E_- = \frac{\gamma_p^2 + E_e^2}{2\gamma_p E_e}}^{\omega -1} \frac{dE_- d\omega dx}{p_-} \frac{n_{\rm ph}(x)}{x^2} W(\omega,E_-,\xi) \label{eq:BHelectron_spec},
\end{align}
where $\epsilon$ is the photon energy in unit of electron rest mass energy.
The kinetic equation for the protons is obtained from consideration of energy conservation
\begin{align}
    \frac{\partial N_{\rm p} (\gamma_p) }{ \partial t } &= \frac{\partial \left ( P_{\rm BH} N_{\rm p} \right )}{\partial \gamma_p}
\end{align}
where the power emitted by Bethe-Heitler is given by 
\begin{align}
    P_{\rm BH} &= \int \gamma_e m_e c^2 \frac{dN_{\rm e^\pm}}{dt} 
\end{align}

\subsection{Photo-hadronic interaction : pion production}

Photo-pion production is the interaction between a proton and a photon mostly producing pions.
This interaction can be divided into four channels : resonance, direct production,
multi-production and fragmentation. In the following, we neglect the contribution
from fragmentation and plan its inclusion for future studies. For the sake of presentation,
in this subsection only, we change the energy unit of the photon distribution from frequency
$\nu$ to energy $\epsilon$ in GeV. Following \cite{HRS10}, the spectral production rate
of each pion species ($\pi^+$, $\pi^-$ and $\pi^0$) can be written as 
\begin{align}
    \frac{d N_{\pi}}{dt } &= \sum_{IT} M_{\rm IT} \int_{\gamma_p} \int_{\epsilon} d\gamma_p d\epsilon  N_p n_{\rm ph} R_{\rm IT} ( \gamma_p , \nu \rightarrow  \gamma_\pi ),
\label{eq:pgamma_pion_spectrum_general}
\end{align}
where the index IT spams all resonances, the two direct production channels, and multi
production channels. In this equation $M^{IT}$ is the multiplicity of each interaction.
This coefficient takes a different value for each interaction and each pion species.
Finally, the rate of interaction is given by \cite{KA09}
\begin{align}
    R_{\rm IT} = \frac{c}{2 \gamma_p^2 \beta_p \epsilon^2} \int_{\rm \epsilon_{\rm th}}^{2\gamma_p \epsilon} d\epsilon_r \epsilon_r \int_\psi d\psi \frac{\partial \sigma_{p\gamma} (\epsilon_r,\psi)}{\partial \psi}  \delta (E_\pi - \xi ) ,
\end{align}
where the threshold energy $\epsilon_{\rm th}$ is constrained by the kinematics of the
reaction, $\epsilon_r$ is the energy of the photon in the frame comoving with the proton,
$\psi$ is the cosine of the comoving (in the proton rest-frame) angle between the photon
direction and the axis representing the direction of the proton in the lab frame, the
differential represent the angular distribution of the reaction and $\xi$ is the pion
energy obtained from the kinematics. For a detailed discussion, on the kinematics, see
\citet{BRS90,BG93a} and \citet{HRS10}.

\subsubsection{Photo-pion production : resonances}

For the cross-section of the nine resonances considered in this work, we consider the
Breit-Wigner approximation 
\begin{align}
    \sigma^{IT} (\epsilon_r) = B^{IT} \frac{s}{\epsilon_r^2} \frac{\sigma_0^{IT} \left ( \Gamma^{IT} \right )^2 s}{ \left ( s- \left (M^{IT} \right )^2 \right )^2 + \left ( \Gamma^{IT}\right )^2s }  ,
\end{align}
where $s = m_p^2 + 2m_p \epsilon_r$ is the total energy in the center of mass. The
parameters $\sigma^{IT}$, $B^{IT}$, $\Gamma^{IT}$ and $M^{IT}$ are  given by
\citet{MER00,HRS10} and in the review of Particle Data Group \citep{PDTG20}. The nine
resonances used are the ones of \citet{MER00}. Contrary to the simple model used by
\cite{HRS10}, we include the angular dependence of the $\Delta$ meson decay for $R_1$ type
resonances, as given by table 3 of \cite{MER00}.

\subsubsection{Photo-pion production : direct production}

For the direct production, we use the parametric cross-sections given by \cite{MER00}
\begin{align}
\sigma_{N\pi} (\epsilon_r) & = 92.7 \mathfrak{P}(\epsilon_r, 0.152,0.25,2) + 40 \exp \left ( - \frac{(\epsilon_r - 0.29)^2}{0.002} \right ) -15\exp \left ( -\frac{(\epsilon_r - 0.37^2)}{0.002}  \right ), \label{eq:pgamma1_tchannel1}\\
\sigma_{\Delta \pi} (\epsilon_r) &=37.7 \mathfrak{P}(\epsilon_r, 0.4, 0.6,2), \label{eq:pgamma1_tchannel2}
\end{align}
where the function $\mathfrak{P}$ is 0 if $\epsilon_r \leq \epsilon_{\rm th}$ and
\begin{align}
\mathfrak{P} (\epsilon_r, \epsilon_{\rm th}, \epsilon_{\rm max}, \alpha) = \left ( \frac{\epsilon_r - \epsilon_{\rm th}}{\epsilon_{\rm max} - \epsilon_{\rm th}}\right )^{A-\alpha} \left (\frac{\epsilon_r}{\epsilon_{\rm max}} \right )^{-A}
\end{align}
otherwise. Here, $A = \alpha \epsilon_{\rm max} / \epsilon_{\rm th}$.
We also include the angular dependence coming from the distribution of the
$t$-Mandelstam variable as explaned in  \cite{MER00}.

\subsubsection{Photo-pion production : multi-pion production}

For the multi-production channel, it is not possible to resort to simple integral expressions. Therefore, we use the approximation developed by \cite{HRS10}. It provides a simple and convenient form for the pion spectrum, while the multiplicities are approximated from results of Sophia \citep{MER00}. 

For completeness, we give here the expressions of the pion spectrum:
\begin{align}
    \frac{\partial N_\pi}{\partial E_\pi} =  \frac{ c m_p}{E_\pi} \sum_{IT} N_p \left( \frac{E_\pi}{\xi^{IT}} \right )  \int_{\epsilon_{\rm th}}^\infty dy n_\gamma \left ( \frac{m_p y \xi^{IT}}{E_\pi} \right ) M^{IT} f^{IT}(y),
\end{align}
where 
\begin{align}
f^{\rm IT} = \left \{ \begin{aligned} 
&0  & & & 2y < \epsilon_{\rm min}^{\rm IT} \\
&\frac{\sigma^{\rm IT}}{(4y^2)}(4y^2 - {\epsilon_{\rm min}^{\rm IT}}^2) & & & ~~~~~ \epsilon_{\rm min}^{\rm IT} \leq 2y < \epsilon_{\rm max}^{\rm IT} \\
&\frac{\sigma^{\rm IT}}{(4y^2)}({\epsilon_{\rm max}^{\rm IT}}^2 - {\epsilon_{\rm min}^{\rm IT}}^2) & & &   \epsilon_{\rm max}^{\rm IT} \leq 2y .
\end{aligned} \right. \label{eq:pg_f_it_multipion_production}
\end{align}
The parameters for the 14 interactions making the approximation are given in Table 6 of \cite{HRS10}.


\subsection{Particle decay}

We consider the decay of charged pions
\begin{align}
\left \{
\begin{aligned}
    \pi^+ &\rightarrow \mu^+ + \nu_\mu, \\
    \pi^- &\rightarrow \mu^- + \bar \nu_\mu ,
\end{aligned} \label{eq:charged_pion_decay_reaction}\right.
\end{align}
of neutral pions
\begin{align}
    \pi^0 \rightarrow 2\gamma,  \label{eq:neutral_pion_decay_reaction}
\end{align}
and of muons
\begin{align}
\left \{
\begin{aligned}
    \mu^- &\rightarrow e^- + \nu_e + \bar \nu_\mu, \\
    \mu^+ &\rightarrow e^+ + \bar \nu_e + \nu_\mu.
\end{aligned}\right.
\end{align}

The kinetic equation of the decaying particle is given by 
\begin{align}
    \frac{\partial N_{\rm i}}{\partial t} &= - \frac{N_{\rm i}}{\tau_{\rm i} \gamma_{\rm i}}  \label{eq:particle_decay_general_kinetic_equation}
\end{align}
where $\tau_{\rm i}$ is the mean-life time of the particle which decays, and $\gamma_{\rm i}$ its Lorentz factor.
The kinetic equation for the daughter particle is given by
\begin{align}
    \frac{\partial N_{\rm j}}{\partial t} \left ( E_{\rm j} \right )= \int_{E_{\rm i} > E_{\rm j}}^\infty \frac{ N_{\rm i}}{\tau_{\rm i} \gamma_{\rm i}} F(E_{\rm i}, E_{\rm j} )dE_{\rm i}, \label{eq:particle_decay_source_general_kinetic_equation}
\end{align}
where $F(E_{\rm i}, E_{\rm j})$ is the spectrum of particle j at energy $E_{\rm j}$ produced by a parent particle of energy $E_{\rm i}$.

The neutrino spectrum from charged pion decay is given by \cite{LLM07}
\begin{align}
    F(E_\pi,E_\nu) = \frac{1}{E_\pi} \frac{1}{1-r_\pi} H\left( 1- r_\pi -\frac{E_\nu}{E_\pi} \right) ,
\end{align}
where $r_\pi = (m_\mu/m_\pi)^2$.

The photon spectrum obtained from neutral pion decay is given by
\begin{align}
    F(E_{\pi_0}, E_\gamma) = \frac{2}{E_{\pi^0}},
\end{align}
where the factor 2 comes from the fact that two photons are created in the decay.

For the electron and neutrino spectra, we do not use the expression given by
\cite{ZK62}, but resort to the simpler relativistic approximation of \cite{LLM07}
\begin{align}
    F_{\nu_e} (E_{\nu_e}, E_\mu) & = \frac{2-6x^2+4x^3}{E_\mu} \label{eq:mu_decay_nue_spec},\\
F_{\nu_\mu}(E_{\nu_\mu}, E_\mu)= F_e(E_e, E_\mu) & = \frac{\frac{5}{3} - 3x^2 + \frac{4}{3}x^3}{E_\mu}, \label{eq:mu_decay_numu_spec}
\end{align}
where $x = E_i/ E_\mu$ for each particle species $i$.

\subsection{Photon and particle escape}

When dealing with a one-zone model, since the emitting region is assumed to be shaped
like a blob, all effects due to photon transport are neglected. Moreover, since we are
considering optically thin plasma, photons cannot accumulate in the emission region for
an arbitrarily large amount of time. Indeed, they would escape the region in which they
are produced in a time of the order of the crossing time
\begin{align}
    t_{\rm ph}^{\rm esc} \sim \frac{2R}{3c}, 
\end{align}
where $R$ is the comoving size. 
In principle, charged particles could remain longer inside the emitting region. Therefore,
particle escape can be treated by adding a term to the kinetic equation of the form
\begin{align}
\frac{\partial N_{\rm i}}{\partial t} &=  \frac{N_{\rm i}}{\mathfrak{T}_i t_{\rm ph}^{\rm esc}} \\
\frac{\partial n_{\rm ph}}{\partial t} &= \frac{n_\gamma}{t_{\rm ph}^{\rm esc}}
\end{align}
where $\mathfrak{T}_i = t_{\rm i}^{\rm esc} /t_{\rm ph}^{\rm esc} $ is a parameter
that is specified for each runs. It represents the time increase it takes for a
particle to escape the system as compared to a photon. In other words, particle
escape is used to crudely represent the finiteness of the emitting region.
In this work, we did not consider particle escape. Instead, we evolve the
distributions until a comoving time equal to the dynamical time scale.

\section{Numerical discretization and prescription in \textit{SOPRANO}}
\label{sec:numerical_discretization}

\numberwithin{equation}{section}

In order to numerically integrate the kinetic equations presented in Section
\ref{sec:kinetic_equation}, a numerical grid for the energy of all particles
is introduced. In this work, \textit{SOPRANO} uses a grid of bins equally space
in logarithmic of the energy\footnote{Note that our numerical method does not
require a uniform grid. Since, it is based on finite volume, we can refine the
grid in one or several energy bands of interest (static mesh refinement). In this
way, we can provide more detailed results in those specific bands, while the rest
of the domain is coarse for faster numerical estimation. This numerical technique
will be used in future works, in which we will study the shape of the spectral peaks.}.
Table \ref{tab:grid_char} gives the grid characteristics for each types of particle,
that is to say the number of energy bins, together with the minimum and maximum energies.
For the energy discretization, we use the approach of the discontinuous Galerkin method,
that we restrain to first order for this paper\footnote{We have implemented some of the
processes with reconstruction up to order 2, but this numerical technique is not included
in the current paper}. On each energy cell $I$, we approximate the distribution
function by a polynomial, while we use for basis the Legendre polynomial basis.
Therefore, on each energy cell $I$, the distribution function is approximated by 
\begin{align}
    N_{\rm i}^I(t, x) = N_{\rm i, 0}^I(t) L_0^I(x)
\end{align}
where the first order Legendre polynomial on the energy cell $I$ is
\begin{align}
    L_0^I = \frac{1}{\sqrt{x_{I+\frac{1}{2}}-x_{I-\frac{1}{2}}}} \equiv \frac{1}{\sqrt{||I||}}
\end{align}
Here $x_{I\pm (1/2)}$ are the energy boundaries of cell $I$ and where we
introduced the additional notation $||I|| = (x_{I+1/2} - x_{I-1/2})$. 
In the following we will use interchangeably $N_i^I \equiv N_{i,0}^I$.

We seek the weak formulation of all kinetic equations presented in Appendix
\ref{sec:kinetic_equation} on each energy interval $I$. For this, we multiply
both sides of any of the kinetic equation by $L_0$ and integrate over $I$. After
simplification, we obtain a system of differential equations for all $N_{\rm i,0}^I$.
This specific discretization and the structure of the kinetic equation allows us
to retrieve a numerical method which conserves energy and the number of particles
when they are conserved. Time discretization is achieved via implicit first order
Euler method. 
leptonic processes, which involved terms of the form $n_i n_j$. 
 Below, we give details on the numerical discretisation on a couple of example and
 give additional details for specific processes when needed.

\begin{table*}
\center
\begin{tabular}{c|c|c|c|}
    Particle & Number of energy cells & Minimum of the grid & Maximum of the grid  \\\hline\hline
    \textit{Photons : } &  150 &  $\nu = 10^{-2}$Hz & $\nu = 10^{30}$ Hz \\\hline
    \textit{Leptons : }&  130 & $\gamma_{\rm e^{\pm}} = 1.2$ &  $\gamma_{\rm e^{\pm}} 5\times 10^{13}$ \\\hline
    \textit{Hadrons : }&  100 & $\gamma_h = 1.2$ & $\gamma_h = 10^{11}$ \\\hline
    \textit{Neutrinos : }& 100 & $E_\nu = 10^{-3}$ GeV &  $E_\nu = 10^{11}$ Gev 
\end{tabular}
\caption{Characteristics of the numerical grids used by \textit{SOPRANO} for the numerical models of this work. The cells are equally space in logarithmic scale.}
\label{tab:grid_char}
\end{table*}

For the sake of the example, consider the synchrotron process and its associated kinetic equations.
Without synchrotron self-absorption, the treatment of synchrotron losses and photon production is
heavily simplified. We start by the photon given by Equation \ref{eq:synch_kinetic_equation_electron}. Multiplying both side by $L_0^J$ and integrating gives
\begin{align}
    \frac{\partial n_{\rm ph}^I}{\partial t} = \frac{1}{h\sqrt{||J||}} \sum_K \frac{N_i^K}{\sqrt{||K||}} \int_K \int_I \frac{j_{\rm synch}^i (\nu,\gamma_i)}{\nu} d\nu d\gamma_i , \label{eq:synch_photon_discretized}
\end{align}
where $N_i$ represent any charged particle. We follow the same procedure for the charge particle
equation and after an integration by part, the kinetic equation can be put into the form
\begin{align}
   & \frac{\partial N_i^K}{\partial t} =  \int_K L_0^K \frac{\partial}{\partial \gamma_k } \left [ P_{\rm synch} N_i^K L_0^K \right ] d\gamma_i \label{eq:synch_kinetic_equation_elec_before_finding_flux}  = \int \frac{\partial L_0^K}{\partial \gamma_i} \left [ P_{\rm synch} N_i^K L_0^K \right ] d\gamma_i -  \frac{1}{\sqrt{K}} \left [ F_{K+\frac{1}{2}} - F_{K-\frac{1}{2}} \right ]  .
\end{align}
In this last expression, only the second term is non-null.
In order to obtain an expression for the numerical fluxes $F_{K+1/2}$ on the right-hand side, we consider the energy lost by particles and that gain by the photons. The total energy gain by the photons is
\begin{align}
    &\frac{\partial E_{\rm ph}}{\partial t} = \frac{\partial }{\partial t} \left ( \sum_J \int_J n_{\rm ph}^J L_0^J h \nu d\nu \right) \label{eq:synchrotron_energy_gain_photon} =  \sum_J \frac{\nu_{J+\frac{1}{2}} - \nu_{J-\frac{1}{2}}}{2 || J ||} \sum_K \frac{N_i^K}{\sqrt{||K||}} \int_K \int_J \frac{j_{\rm synch} (\nu,\gamma_i)}{\nu} d\gamma_i d\nu   ,  
\end{align}
where the last equality is obtained after using Equation \ref{eq:synch_photon_discretized}.
Turning to the energy lost by the charged particles, we have
\begin{align}
    \frac{\partial E_i}{\partial t} = m_i c^2 \sum_K \frac{\partial N_i^K}{\partial t} \frac{\gamma_{i,K+\frac{1}{1}}^2 - \gamma_{i,K-\frac{1}{1}}^2}{2 \sqrt{||K||}} .   
\end{align}
Inserting the expression for the time derivative of the distribution function coefficients,
and reorganising the summation it comes
\begin{align}
    &\frac{\partial E_i}{\partial t} = m_i c^2 \left [ - \frac{\gamma^2_{i,\frac{1}{2}} - \gamma^2_{i,-\frac{1}{2}}}{2(\gamma_{i,\frac{1}{2}} - \gamma_{i,-\frac{1}{2}})} F_{-\frac{1}{2}} +  \frac{\gamma^2_{i,\xi+\frac{1}{2}} - \gamma^2_{i,\xi-\frac{1}{2}}}{2(\gamma_{i,\xi+\frac{1}{2}} - \gamma_{i,\xi-\frac{1}{2}})} F_{\xi + \frac{1}{2}} + \sum_K \left ( \frac{\gamma^2_{i,K+\frac{1}{2}} - \gamma^2_{i,K-\frac{1}{2}}}{(\gamma_{i,K+\frac{1}{2}} - \gamma_{i,K-\frac{1}{2}})} - \frac{\gamma^2_{i,K+\frac{3}{2}} - \gamma^2_{i,K+\frac{1}{2}}}{(\gamma_{i,K+\frac{3}{2}} - \gamma_{i,K+\frac{1}{2}})} \right ) \frac{F_{K+\frac{1}{2}}}{2} \right ]   
\end{align}
where in this specific equation $\xi$ represent the number of cells of the grid for
particle species $i$. We assume that no particle diffuse out of the energy grid.
This gives $F_\xi = F_{-\frac{1}{2}} = 0$. Inverting the summation order in Equation
\ref{eq:synchrotron_energy_gain_photon}, and identifying the term gives the expression
of the fluxes
\begin{align}
& \mathcal{F}_{K+\frac{1}{2}} = \frac{1}{m_i c^2} \frac{ N_{i,0}^{K+1} / \sqrt{\gamma_{i, K+\frac{3}{2} } - \gamma_{i, K+\frac{1}{2} }}}{\left [ \frac{\gamma_{i, K+\frac{1}{2}}^2-\gamma_{i, K-\frac{1}{2}}^2}{\gamma_{i, K+\frac{1}{2} } - \gamma_{i, K-\frac{1}{2} }} \right ]  - \left [ \frac{\gamma_{i, K+3/2}^2-\gamma_{i, K+\frac{1}{2}}^2}{\gamma_{i, K+\frac{3}{2} } - \gamma_{i, K+\frac{1}{2} }} \right ] }  \times  \sum_J  \frac{\nu_{J+\frac{1}{2}}^2 - \nu_{J-\frac{1}{2}}^2}{\nu_{J+\frac{1}{2}} - \nu_{J-\frac{1}{2}}} \int_J \int_{K+1}   \frac{j_{synch} (\nu,\gamma_i)}{\nu}  d\gamma_i  d\nu.
\end{align}
This choice of the flux preserves the total energy of the system, while the particle number
conservation is ensured by the structure of Equation
\ref{eq:synch_kinetic_equation_elec_before_finding_flux}. However, this choice of numerical
discretization leads to numerical diffusion, see \ref{sec:Test_sync_CC}, in which the current
scheme is compared to the classical Chang and Cooper scheme \citep{CC70}.

\ifx\ExtendedAppendix\undefined
For the Compton scattering process, the weak formulation is trivially obtained and does
not require the use of integration by part, since the kinetic equation does not take the
form of a diffusion equation. For all other processes, their respective kinetic equation
takes either the same form as that of synchrotron mechanism or of Compton scattering.
Therefore, all energy discretization can be easily obtained following the same procedure
outlined above. We note, that we use redistribution of particles in integro-differential
equation type to preserve simultaneously total particle number (when required by the process)
and total energy.

\else
\subsection{Energy discretization of inverse Compton scattering process}
We start by the kinetic equation of the photons given by Equation \ref{eq:kinetic_IC_photon}. For each energy cell $I$, we multiply by $L_0$ and integrate over $I$ to obtain after simplification
\begin{align}
\frac{\partial n^{J}_{\rm ph}}{\partial t} &= \frac{1}{\sqrt{||J||}} \sum_{k}  \sum_{I < J} \frac{N_{\rm e}^k}{\sqrt{||K||}} \frac{n_{\rm ph}^{I}}{\sqrt{||I||}}\sigma_{IKJ} \nonumber \\
-& \frac{ n_{\rm ph}^{J}}{||J||} \sum_{k} \frac{N_{\rm e}^{k}}{\sqrt{||K||}} \sum_{I>J} \sigma_{JKI} 
\label{eq:CS_photon_discrete}
\end{align}
where the condition on $I$ and $J$ come from the ultra-relativistic approximation of the Jones kernel, \textit{i.e.} photon do not lose energy when they interact with ultra-relativistic electrons. In addition
\begin{align}
    \sigma_{IKJ} \equiv \int_I\int_J\int_K \sigma(\nu_I,\gamma_K \rightarrow \nu_J) d\nu_I d\nu_J d\gamma_K
\end{align}
was tabulated on each nuple $(I,J,K)$ by Gauss-Legendre integration. It is clear that this discretization preserves total number to machine accuracy.

We now turn to the kinetic equation for electron positron \ref{eq:ke_electron_CS}. Multiplying both side by the Legendre function, developing the expression of the distribution function and integrating both side of the equation (with an integration by part on the right-hand side) we finally obtain
\begin{align}
    \frac{\partial N_e^K}{\partial t} &= \left [L_0^K P_C N_e^K \right ] - \int_K \frac{\partial L_0^K}{\partial \gamma_e} P_c N_e^K d\gamma_e \nonumber \\
    &= \frac{1}{\sqrt{||K||}} \left [ F_{K+\frac{1}{2} } - F_{K_\frac{1}{2}} \right ]  \label{eq:CS_electron_equation_with_fluxes}
\end{align}
We could use this expression directly with the value of the power radiated by Compton scattering, \cite{RL79}. However this approach would not guaranty energy conservation. Therefore, in order to obtain an expression for the numerical fluxes, we consider conservation of energy of the Compton scattering process : the energy loss by the electrons is exactly gained by the photons.

The variation of energy of the electron is
\begin{align}
\frac{\partial E_e}{\partial t} &=  \frac{\partial }{\partial t} \left ( \sum_K \int_K N_e^K L_0^K \gamma_e m_e c^2 d\gamma_e \right ) \\
&= m_e c^2 \sum_K \frac{\partial N_e^K}{\partial t} \frac{\gamma_{e,K+\frac{1}{2}}^2 - \gamma_{e,K-\frac{1}{2}}^2}{2\sqrt{||K||}} \\
&= m_e c^2 \sum_K \left [F_{K+\frac{1}{2}} - F_{K-\frac{1}{2}} \right ]  \frac{\gamma_{e,K+\frac{1}{2}}^2 - \gamma_{e,K-\frac{1}{2}}^2}{2||K||}
\end{align}
where the last equation is obtained by using Equation \ref{eq:CS_electron_equation_with_fluxes}. We reorder the equation to obtain
\begin{align}
&\frac{1}{m_e c^2} \frac{\partial E_e}{\partial t} = \nonumber \\
& \sum_K \left (  \frac{\gamma_{e,K+\frac{1}{2}}^2 - \gamma_{e,K-\frac{1}{2}}^2}{2||K||} - \frac{\gamma_{e,K+3/2}^2 - \gamma_{e,K+\frac{1}{2}}^2}{2||K+1||}\right ) F_{K+\frac{1}{2}}  
\end{align}
We now write the variation of photon energy
\begin{align}
\frac{\partial E_{\rm ph}}{\partial t} =  h \sum_J \frac{\partial n_{\rm ph}^J}{\partial t} \frac{\nu_{J+\frac{1}{2}}^2 - \nu_{J-\frac{1}{2}}^2}{2\sqrt{||J||}}
\end{align}
in which we can use Equation \ref{eq:CS_photon_discrete}. 
Identification of the terms for a specific electron index $K$ gives the definition of the fluxes. After simplification, they can be writen as 
\begin{align}
&F_{K+\frac{1}{2}} =   \frac{N_e^{K+1}}{\sqrt{||K+1||}} \frac{\frac{h}{m_e c^2}}{ \left (  \frac{\gamma_{e,K+\frac{1}{2}}^2 - \gamma_{e,K-\frac{1}{2}}^2}{||K||} - \frac{\gamma_{e,K+3/2}^2 - \gamma_{e,K+\frac{1}{2}}^2}{||K+1||}\right)} \nonumber \\
&\times \sum_J \frac{\nu_{J+\frac{1}{2}}^2 - \nu_{J-\frac{1}{2}}^2}{\sqrt{||J||}} \left [ \sum_{I} \frac{ n_{\rm ph}^{I}\sigma_{I(K+1)J}}{\sqrt{||I||||J||}}   -\frac{ n_{\rm ph}^{J}   }{||J||} \sum_{I} \sigma_{J(K+1)I}  \right ]  \label{eq:CS_electron_fluxes}
\end{align}
Using this expression ensures number conservation, as well as energy conservation by construction of the discrete equation. However it leads to important numerical diffusion, see \ref{sec:Test_sync_CC}.

\subsection{Energy discretization of pair creation process}

\subsection{Energy discretization of Bethe-Heitler pair production and associated proton cooling}

For the Bethe-Heitler process, we compute the rate at which pairs are produced from integrating the spectral rate over the proton distribution function. 
For the cooling of the protons, we proceed similarly to synchrotron cooling such that the final expression for the numerical flux for the proton is 
\begin{align}
&F_{J+\frac{1}{2}} = \frac{N_p^{J+1}}{  \frac{\gamma_{p,J+\frac{1}{2}}^2 - \gamma_{p,J-\frac{1}{2}}^2}{||J||} - \frac{\gamma_{p,J+3/2}^2 - \gamma_{p,J+\frac{1}{2}}^2}{||J+1||} }  \\
&\times \left [2 \frac{m_e}{m_p} \sum_K \sum_I \frac{\gamma_{e,K+\frac{1}{2}}^2 - \gamma_{e,K-\frac{1}{2}}^2}{||K||}  \frac{ N_{ph}^{I}}{\sqrt{||I||||J+1||}} \sigma_{IJ+1\rightarrow K}^{BHs} \right ] \nonumber
\end{align}
where
\begin{align}
&\sigma(IJ\rightarrow K) \\
=& \int_K  dE_e   \int_{J } dE_p   \int_{I }^\infty \int_{ \omega_{\rm min} }^{2\gamma_p \epsilon} \int_{E_{-, \rm min} }^{\omega -1} \frac{dE_- d\omega d\epsilon}{2 p_-\epsilon^2 \gamma_p^3}  W(\omega,E_-,\xi) \label{eq:BH_electron_source_discretized} \nonumber 
\end{align}
where $\epsilon_{\rm min} = (\gamma_p + E_e)^2/(4\gamma_p^2 E_e)$, $\omega_{\rm min} = (\gamma_p + E_e)^2/(2\gamma_p E_e)$, $E_{-, \rm min} = (\gamma_p^2 + E_e^2)/(2\gamma_p E_e)$, index $J$ is such that $E_p > E_e$ and finally index I is such that $\epsilon_{I-\frac{1}{2}} > \epsilon_{\rm min}$.

\subsection{Energy discretization of Photo-pion interaction}

Similarly to the Bethe-Heitler process, the pion spectrum is obtained by integrating the spectral rate over proton and photon distribution function. Once the integrals over all energy cells are computed, the pion kinetic energy is reduced to a trivial source term in all pion energy bins. Specifically in this section, the photon spectrum is expressed in unit of GeV.

\subsubsection{Resonances}

\subsubsection{Direct production}

\subsubsection{Multi-production}

\subsubsection{Cooling and sink term}

The only difficult resides in the cooling of protons and in the transformation of a proton into a neutron. We split all interactions based on their final product (proton and neutron) and take them into account in a different way.

\subsection{Energy discretization of two particle decay}
\label{sec:two_particle_decay_discreet_kinetic_equation}

\authorcomment{We need to put proper units here}

In our analysis the particles that decay to form two different particles are charged and neutral pions given by Equations \ref{eq:charged_pion_decay_reaction} and  \ref{eq:neutral_pion_decay_reaction}. In order to obtain a conservative scheme, we rewrite the kinetic equation \ref{eq:particle_decay_source_general_kinetic_equation} for muon and neutrino in the form
\begin{align}
    \frac{\partial N_{ \mu}}{\partial t} &= \int_0^\infty \frac{N_\pi}{\tau_\pi \gamma_\pi} \int_0^{E_\pi} F(E_\nu) \delta( E_\pi - E_\mu - E_\nu) dE_\pi dE_\nu \\
    \frac{\partial N_{ \nu}}{\partial t} &= \int_0^\infty \frac{N_\pi}{\tau_\pi \gamma_\pi} \int_0^{E_\pi} F(E_\nu) \delta( E_\pi - E_\mu - E_\nu) dE_\pi dE_\mu 
\end{align}
while the kinetic equation for pions is written 
\begin{align}
    &\frac{\partial N_{ \mu}}{\partial t} \\
    =& -\int_0^\infty \frac{N_\pi}{\tau_\pi \gamma_\pi} \int_0^{E_\pi} \int_0^{E_\pi} F(E_\nu) \delta( E_\pi - E_\mu - E_\nu) dE_\mu dE_\nu \nonumber
\end{align}
Following the same procedure, we multiply both side of the equation by the basis function and integrate on each interval to obtain
\begin{align}
    ||I|| \frac{\partial N_\pi^I}{\partial t} &=  -\frac{N_\pi^I}{\tau_\pi } \sum_{J<I} \sum_{K<I} \sigma_{\dot \pi^\pm}(I,J,K) \\
    \sqrt{||K||} \frac{\partial N_\mu^K}{\partial t}  &=  \sum_{I > K} \frac{N_\pi^I }{\tau_\pi \sqrt{||I||}} \sum_{J<I}  \sigma_{\dot \pi^\pm}(I,J,K) \\ 
    \sqrt{||J||} \frac{\partial N_\nu^J}{\partial t} &=  \sum_{I > K} \frac{N_\pi^I}{\tau_\pi  \sqrt{||I||}} \sum_{K<I} \sigma_{\dot \pi^\pm}(I,J,K)
\end{align}
where the cross-section of pion decay is
\begin{align}
    \sigma_{\dot \pi^\pm}(I,J,K) = \int_J \int_I \int_K  \frac{F(E_\nu)}{ \gamma_\pi}  \delta (E_\pi - E_\mu - E_\nu) dE_\pi dE_\mu dE_\nu
\end{align}
The triple integrals are pre-computed and tabulated on the grids. Summation over all cells shows that the number of pion that decay is equal to machine accuracy to the number of created muon and neutrino. However, energy is not conserved to machine accuracy. To achieve this we resort to the procedure describe in Appendix \ref{sec:energy_conservation_scheme}. 

The procedure is the same for $\pi^0$ decay although the equations are simplified since only photons are created. Following the same methodology, we cast the discreet photon kinetic equation in the following form
\begin{align}
    \frac{\partial n_{\gamma}^J}{\partial t} &=  \frac{2}{\tau_{\pi_0} } \frac{1}{\sqrt{||J||}} \sum_K \frac{N_{\pi_0}^K}{\sqrt{||K||}} \sigma_{\dot \pi^0}
\end{align}
while the discreet kinetic equation for the neutral pion is obtained as
\begin{align}
        \frac{\partial N_{\pi_0}^K}{\partial t} &=  \frac{N_{\pi_0}^K}{\tau_{\pi_0} ||K|| } \sum_{J < K} \sigma_{\dot \pi^0}
\end{align}
where
\begin{align}
\sigma_{\dot \pi^0} = \int_J  dE_{\gamma} \int_{{\rm max} \left (E_\gamma , E_{\pi_0, K-\frac{1}{2}}\right )}^{E_{\pi_0, K+1}} \frac{dE_{\pi_0}}{\gamma_{\pi_0} E_{\pi_0}}
\end{align}
Energy is also conserved following the procedure given in Appendix \ref{sec:energy_conservation_scheme}.

\subsection{Energy discretization of three particle decay}

The process is basically the same as that given in Section \ref{sec:two_particle_decay_discreet_kinetic_equation}. The difference is that it is not possible to obtain a simple conservative differential equation by introducing $\delta-$function in the kinetic equation. To achieve number conservation of all species, we resort to normalization of the cross-section such that the number of created particle of all species is equal. We note that for all discreet cross-section the normalization is smaller than one percent. For the reference particle, we (arbitrarily) chose the electrons/positrons, and re-normalize the number of created neutrinos to the number of created electrons/positrons.

Following the same methodology, we obtain the kinetic equation for muon and electrons/positrons
\begin{align}
    \frac{\partial N_{\mu}^{I}}{\partial  t}  & = - \frac{N_{\mu}^{I}} {\tau_\mu ||I||}\sum_{K<I} \int_{K} \int_{I} \frac{1}{\gamma_\mu}  F_e (E_e, E_\mu) dE_edE_\mu\\
    \frac{\partial  N_{e}^{K}}{\partial  t}  & = \frac{1}{\tau_\mu} \sum_{I > K} \frac{N_{\mu}^{I}}{\sqrt{||K||||I||}} \int_{K}\int_{I} \frac{F_e (E_e, E_\mu)}{\gamma_\mu}  dE_ed E_\mu
\end{align}
while the discreet kinetic equation for both neutrino species are given by
\begin{align}
    \frac{dN_{\nu_e}^{J}}{dt}   & = \frac{1}{\tau_\mu} \sum_{I >J} N_{\mu}^{I} \int_{J}\int_{I} \frac{F_{\nu_e} (E_{\nu_e}, E_\mu)  }{\gamma_\mu \sqrt{||I||||J||}}   dE_{\nu_e}dE_\mu \label{eq:muon_decay_neutrino_nu_e_discreet}\\
    \frac{dn_{\nu_\mu}^{M}}{dt}  & = \frac{1}{\tau_\mu} \sum_{I>M} N_{\mu}^{I} \int_{M}\int_{I} \frac{F_{\nu_\mu} (E_{\nu_\mu}, E_\mu)}{\gamma_\mu\sqrt{||I||||M||}}   dE_{\nu_\mu}dE_\mu \label{eq:muon_decay_neutrino_nu_mu_discreet}
\end{align}

For a given muon energy bin I, the total rate at which a particle $i$ are created is given by
\begin{align}
    \dot N_{i}^I = \sum_K \int_K \frac{\partial N_i^K}{\partial t} L_0^K d\gamma_e = \frac{N_{\mu}^{I}}{\tau_\mu\sqrt{||I||}} \sum_K   R_i(K,I)
\end{align}
where 
\begin{align}
    R_i(K,I) = \int_{K}\int_{I} \frac{F_i (E_i, E_\mu)}{\gamma_\mu}  dE_i dE_\mu
\end{align}
We enforce the number of created neutrino of both species independently to be equal to the number of created electrons. For this we multiply Equation \ref{eq:muon_decay_neutrino_nu_e_discreet} and \ref{eq:muon_decay_neutrino_nu_mu_discreet} by the expected number of particle created to the actual number of particle created, that is to say we multiply the electronic neutrino equation by the coefficient
\begin{align}
    A_{\nu_e} = \frac{\sum_{K <I} R_e(K,I)}{\sum_{J<I} R_{\nu_e} (J,I)}
\end{align}
while the antineutrino muonique equation is multiplied by 
\begin{align}
    A_{\nu_\mu} = \frac{\sum_{K <I} R_e(K,I)}{\sum_{J<I} R_{\nu_\mu} (J,I)}
\end{align}
Once the rate have been corrected, we resort to the scheme described in Appendix \ref{sec:energy_conservation_scheme} to preserve energy.

\fi


Before discussing the temporal discretization, we note that the equation are non-linear
in the distribution function for Compton scattering, pair production, photo-pion and
photo-pair processes. We decided to linearize the kinetic equations of photo-pion and
photo-pair processes by assuming that the target photon-field is equal to the one at the
previous time step, effectively making those process linear in the distribution functions.
For all leptonic processes, we preserve the non-linearity of the kinetic equations and
solve at each time step a non-linear system via the Newton-Raphson method. Since the
gradients can be computed analytically, we do not need to use numerical estimates for
the Jacobian. The temporal evolution of the distribution function is performed with
the first order implicit Euler method.

One time step of the code takes the following form
\begin{enumerate}
\item solve the linear kinetic equation to obtain the protons and neutrons spectra at
time $t + dt$, assuming that the photon distribution function is given at time $t$ for 
photo-pion and photo-pair processes. The pairs and photons created in those two processes
and by the proton synchrotron process are saved to be use as a source term in the leptonic
computation.
\item compute the decay and cooling (when required) of pions and muons. The pairs and
photons created in the muon and pion decay, as well as their synchrotron radiation are
saved to be used as source terms in the leptonic computation.
\item perform the non-linear implicit leptonic computation with the source terms computed
in the two previous steps.
\end{enumerate}

\section{Code tests and examples}

\label{sec:code_tests}

\subsection{Synchrotron and inverse Compton cooling for electrons}
\label{sec:Test_sync_CC}

\begin{figure}
    \centering
    \begin{tabular}{cc}
    \includegraphics[width=0.45\textwidth]{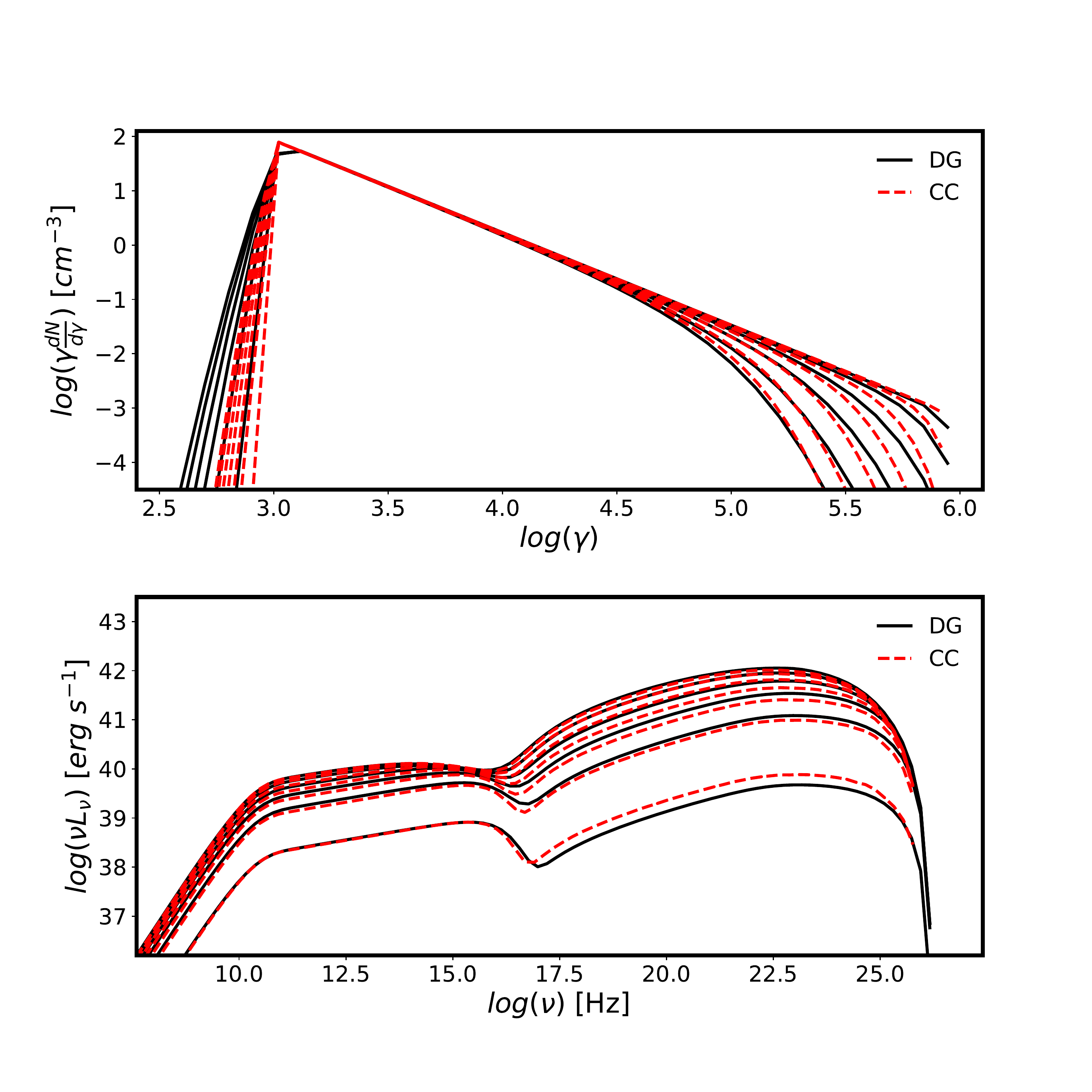} &
    \includegraphics[width=0.45\textwidth]{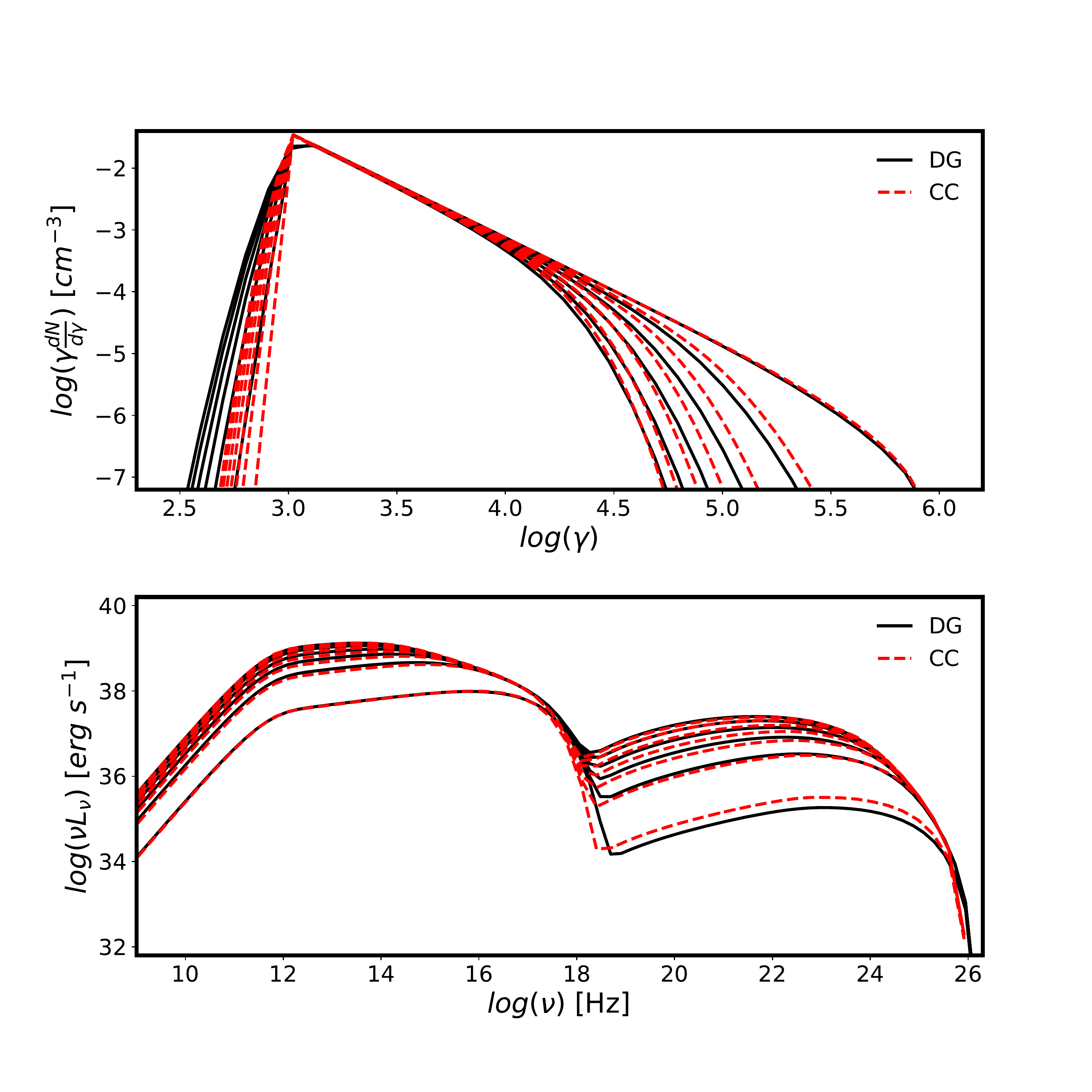}
    \end{tabular}
    \caption{ Comparison between the \citet{CC70} scheme and \textit{SOPRANO} for Compton
    scattering and synchrotron radiation in two different regimes, whose parameters are given
    in Table \ref{tab:appendixC_synch}. Left $-$ inverse Compton dominates. Right $-$ synchrotron
    cooling dominates. The plots show the time evolution of the electron (top) and photon
    (bottom) distribution functions up to dynamical time scale, in the comoving frame. Electron number conservation
    is satisfied to machine accuracy ($10^{-16}$) while the energy is conserved to an accuracy
    better than $10^{-11}$.}
    \label{fig:DGvsCC}
\end{figure}

\begin{table*}
    \centering
   	\begin{tabular}{@{}l c c}
 		\hline & 
		  IC dominance & Syn dominance \\
 		\hline
 $R/10^{17}\:{\rm cm}$ & 1 & 1 \\
 $B[G]$ & 0.005 & 0.1 \\
 $\gamma_{\rm e, min} $ & $10^3$ & $10^3$ \\
 $\gamma_{\rm e, max}$ & $9\times10^5$ & $9\times10^5$\\
 $\alpha_{\rm e}$  & 2.7 & 2.7 \\
 $U_{\rm e}/U_{\rm B}$& $10^{5}$ &$0.1$\\

 		\hline
 		\end{tabular}
 	\caption{Parameters used for our numerical comparison between \textit{SOPRANO} and the Chang and Cooper scheme \protect\cite{CC70}. The corresponding evolution of the photon and electron distribution functions is given on Figure \protect\ref{fig:DGvsCC}.}
 	 \label{tab:appendixC_synch}
 		\end{table*}


In this section, we present two tests performed for the synchrotron and inverse Compton radiation
processes. First, the results from \textit{SOPRANO} are compared to the results obtain with our
implementation of the \cite{CC70} scheme, which is widely spread and used in time-dependent
application, see \textit{e.g.} \citet{CG99,GPW17}. We consider a situation in which electrons
are continuously injected into the radiating zone in the form of a power-law between
$\gamma_{\rm min} = 10^3$ and $\gamma_{\rm max} = 9\times 10^5$. The properties of the radiating zone
are such that for one test, inverse Compton cooling dominates over synchrotron cooling, while for
the other test it is the opposite. The parameter are summarized in Table \ref{tab:appendixC_synch}.

Figure \ref{fig:DGvsCC} shows the results. It is clear that the agreement between the Chang and Cooper
scheme and \textit{SOPRANO} is very good for both the electron and the photon distribution functions.
We notice that \textit{SOPRANO} is more diffusive at low energy below the peak of the electron
distribution function. This is mostly because of the scheme used to preserve the total
energy of the system in \textit{SOPRANO}, which induces extra diffusion. However, despite
these differences for the electron distribution function, the photon spectra are in very good agreement. 






\subsection{Proton cooling on black-body photons by photo-pair and photo-pion interaction}

\begin{figure}
    \centering
    \begin{tabular}{cc}
    \includegraphics[width=0.45\textwidth]{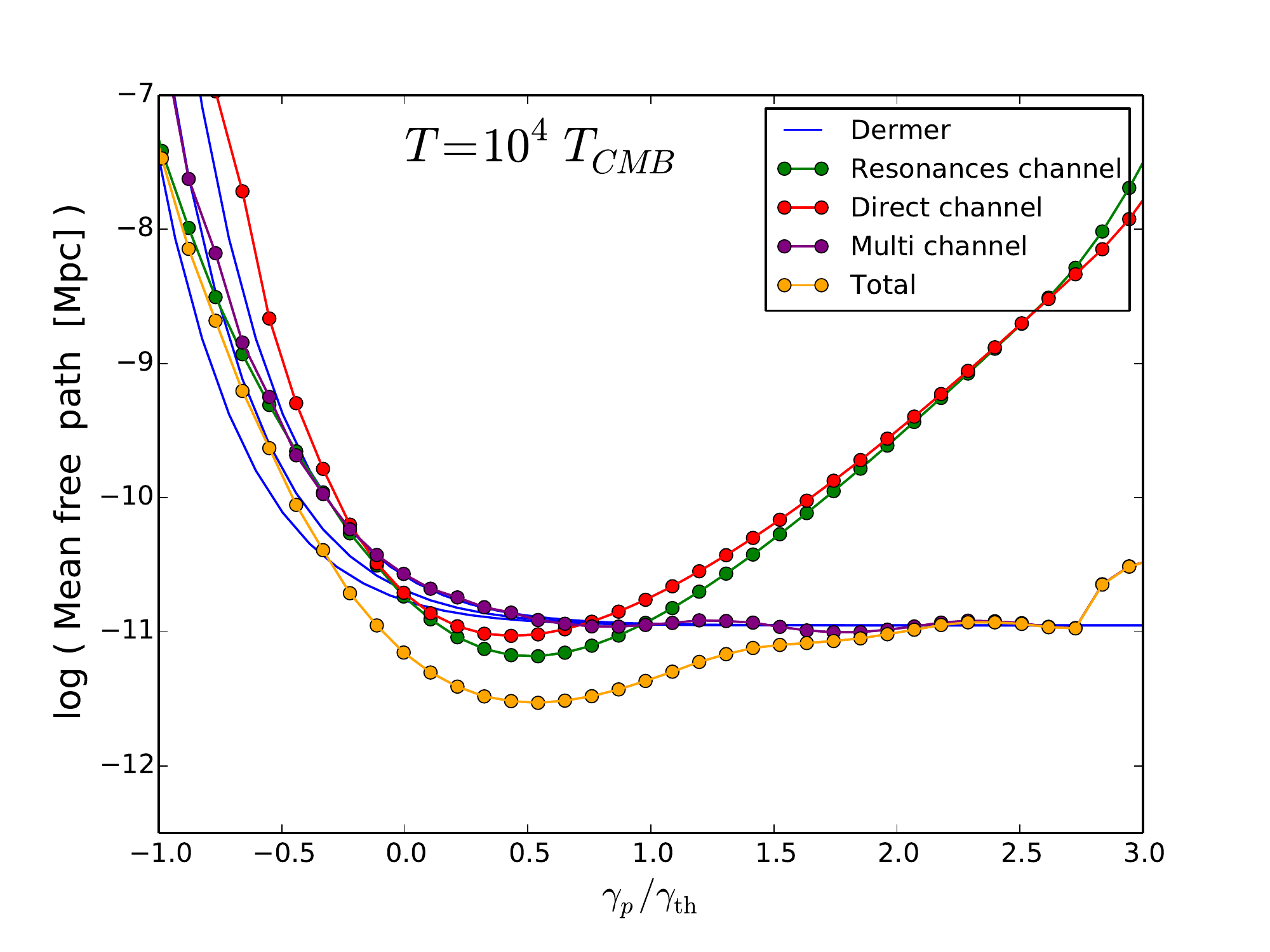} &
    \includegraphics[width=0.45\textwidth]{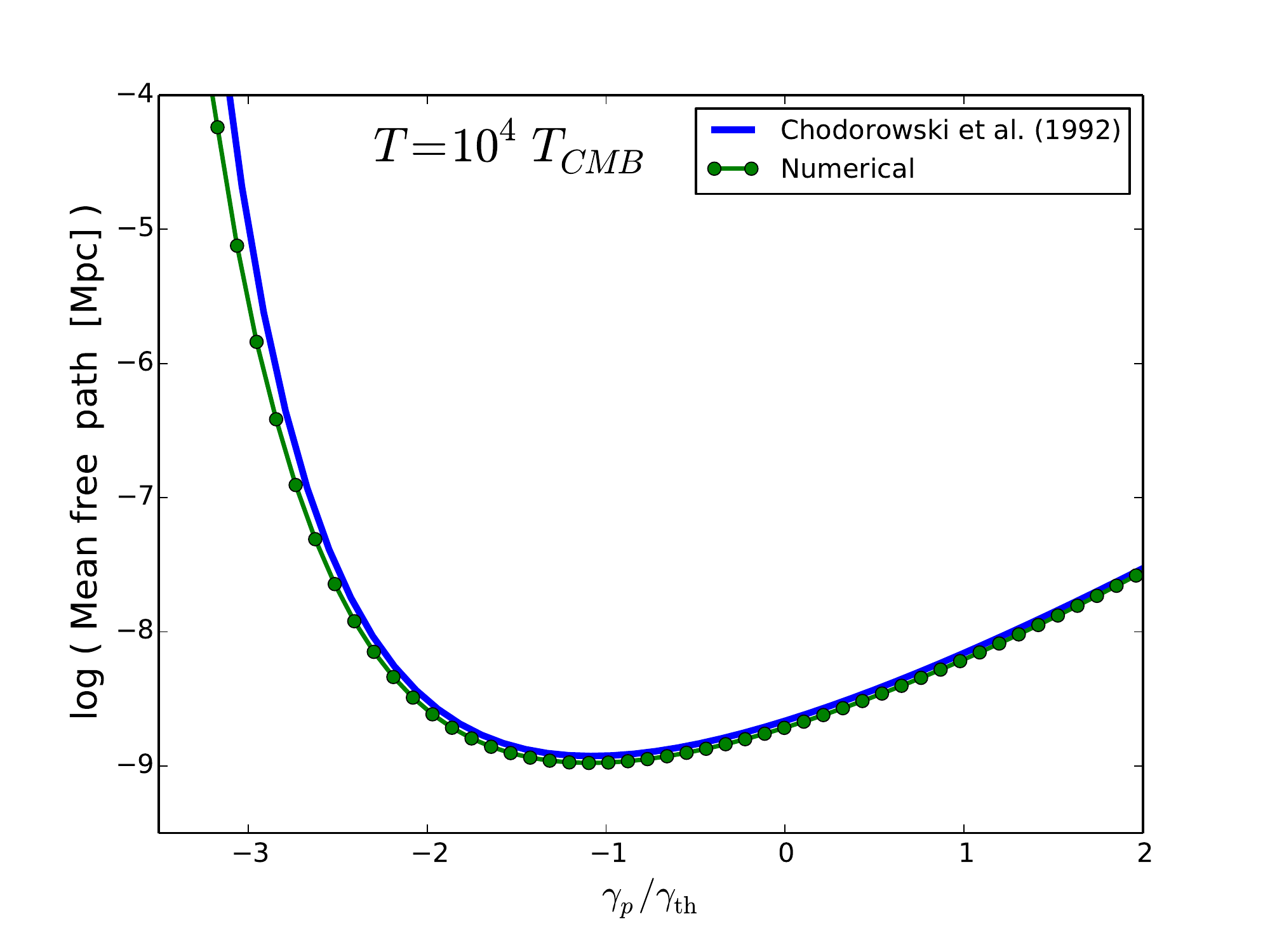}
    \end{tabular}
    \caption{ Right : Proton mean free path as a function of their Lorentz factor for the
    photon-pion process. The photon temperature is $T = 10^4T_{\rm CMB}$. Blue : mean free
    path computed by using Equation \ref{eq:Dermer_cooling_time_photopion} for the lower boundary
    center and upper boundary of an energy cell. The Lorentz factor is normalized by
    $\gamma_{\rm th} = \epsilon_{\rm th} / (2k_{\rm B} \theta) \sim 3.2\times 10^7 $.
    Right : Proton mean free path as a function of the Lorentz factor for the photo-pair process.
    Green - \textit{SOPRANO}. Blue with the approximation of  \citet{CZS92}. For easy comparison,
    the proton Lorentz factor is also normalized by $\gamma_{\rm th}$.}
    \label{fig:Dermer_comp}
\end{figure}

We start by computing the proton cooling time in photo-pion production.
\cite{AD03} presented a simple model for the cross-section and inelasticity
for the photon-pion interaction. The model is such that the product of the
cross-section with the inelasticity is constant for all energies larger than
the threshold energy in the center of mass frame. It gives a simple estimate
of the cooling time for a proton of Lorentz factor $\gamma_p$ interacting with
an isotropic photon field with the spectrum of a black-body withe temperature
$\theta$. From \cite{DM09}, it reads
\begin{align}
t_{p\gamma}^{-1} (\gamma_p)= \frac{8\pi c \sigma K \theta^3}{\lambda_c^3} \int_\omega ^ \infty dy \frac{y^2-\omega^2}{\exp(y) - 1}, \label{eq:Dermer_cooling_time_photopion}
\end{align}
where $\lambda_c$ is the Compton wavelength, $\sigma K \sim 70 \mu$b and 
\begin{align}
    \omega = \frac{\epsilon_{\rm th}}{2 \gamma_p \theta}.
\end{align}
In Figure \ref{fig:Dermer_comp}, we present the mean free path in Mpc as a function
of proton Lorentz factor for an hypothetical radiation field with temperature
$10^4 T_{\rm CMB}$, with the temperature of the cosmic microwave background
$T_{\rm CMB} = 2.725$K. To obtain this plot, we considered a $\delta$-function in each
of the proton energy grid bins. We see that the agreement is quite good. We remark that the
contribution of the multi-production channel has a sharp increase at large Lorentz factor.
This is not physical and is a grid effect. Indeed, the proton cooling is computed by summation
over all created pions. Because both the pion and proton energy grid have the same maximum
Lorentz factor, protons at the highest energies in our grid do not interact substantially,
since they cannot create pions of the correct energy.

We now turn to the photo-pair process. The energy loss rate for this process can be written as
\begin{align}
    \frac{dE}{dt} = \alpha_f r_e^2 c m_e c^2 \int_2^{\infty} d\epsilon n_{\rm ph} \left ( \frac{\epsilon}{2\gamma_p}\right ) \frac{\phi(\epsilon}{\epsilon^2}  .
\end{align}
\cite{CZS92} gives a convenient approximation for the differential cross-section $\phi(\epsilon)$, see their appendix. It is therefore easy to compute the energy loss pathlength $r = c [(dE/dt)/E]^{-1}$. A comparison between this semi-analytical approach and our numerical discretization is given by the right part of Figure \ref{fig:Dermer_comp}. We see that the agreement is excellent.


\subsection{Decay time}

\begin{figure}
    \centering
    \begin{tabular}{cc}
    \includegraphics[width = 0.75 \textwidth]{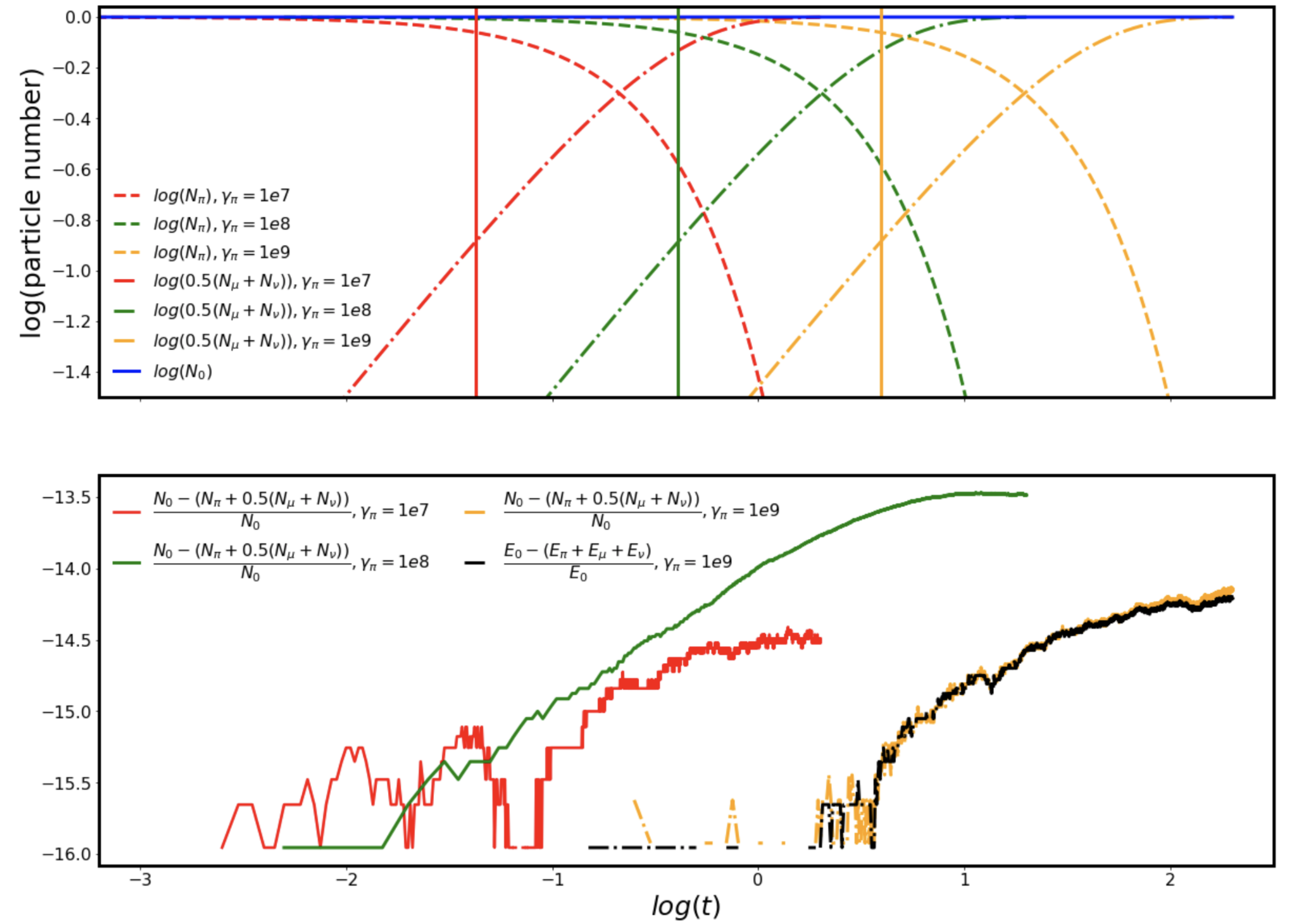}
    \end{tabular}
    \caption{Top - evolution of the pion, neutrino and muon numbers as a function of time for
    the pion decay process. The vertical lines correspond to the decay time of the corresponding
    particle Lorentz factor. Obviously the decay time is properly respected. Bottom - temporal
    evolution of the particle number conservation and energy conservation. The total energy
    conservation is only shown for $\gamma_pi = 10^9$, but similar results are obtained for
    other particle energy. }
    \label{fig:pi_decay_test}
\end{figure}

We show in this subsection how particles decay in our code with the examples of pion decay.
We inject pions with a specific Lorentz factor and simulate the evolution of the system as
they decay, producing neutrinos and muons. The initial pions Lorentz factor are
$\gamma_\pi = 10^7, ~ 10^8,~ 10^9$. In this section only, we assume that muons cannot decay.
Figure \ref{fig:pi_decay_test} shows the evolution of the pion, neutrino and muon numbers as
time evolve. The muon and neutrino number is obtained by summing over their respective
distribution function. The same figure also shows the evolution of particle number and
total energy of the system, which can be seen to be satisfied to accuracy better than
$10^{-13}$ after $4 \times 10^3$ iterations.

\twocolumn

\end{document}